\documentclass[letterpaper, oneside, 12pt]{article}
\usepackage{amssymb,amsmath,graphicx,epsfig,amsfonts,verbatim,natbib}

\usepackage[top=1.0in, bottom=1.25in, left=1.0in, right=1.0in]{geometry}
\usepackage[onehalfspacing]{setspace}
\usepackage{rotating,longtable,booktabs, colortbl}       % needed for rotating
\usepackage{caption}
\usepackage{mdwlist}
\usepackage{authblk}    % to add author affiliation
\usepackage[shortlabels]{enumitem}
\usepackage{lscape,afterpage}                    % supplies a landscape
\usepackage{appendix}
\usepackage{url,hyperref}                                            % for emails. Also, it links all your references...
\usepackage{xcolor}
\usepackage{soul}
\usepackage[normalem]{ulem}
\usepackage{siunitx} % For aligning numbers by decimal point
\usepackage[standard]{ntheorem}
\usepackage[algo2e]{algorithm2e}

\theoremstyle{break}
\theorembodyfont{\normalfont}
\newtheorem{algorithm}[algocf]{Algorithm}

\begin{document}
%\title{Does Accounting for Misclassification Alter Determinants of Reporting Spousal Violence? Evidence from Bayesian Quantile Regression}

%\title{Modeling Misclassification in Spousal Violence Reporting: Evidence from Bayesian Quantile Regression}
\title{Bayesian Quantile Regression for Misclassified Binary Data with an Application to Spousal Violence Reporting}
%\title{Bayesian Quantile Regression for Misclassified Binary Data}

\author[1]{Joon Jin Song\thanks{Corresponding author. E-mail address: Joon\_Song@baylor.edu (Joon Jin Song)} }
\author[2]{Mohammad Arshad Rahman}%\thanks{marshad@iitk.ac.in}}
\author[3]{Yoo-Mi Chin}%\thanks{Yoo-Mi\_Chin@baylor.edu}}
\author[1]{James Stamey}%\thanks{James\_Stamey@baylor.edu}}

\affil[1]{Department of Statistical Science, Baylor University, Waco, Texas, USA}
\affil[2]{Department of Economic Sciences, Indian Institute of Technology Kanpur, Kanpur, IN}
\affil[3]{Department of Economics, Baylor University, Waco, Texas, USA}

\smallskip
\date{\today}
\maketitle
%
%
%%%%%%%%%%%%%%%%%%%%%%%%%%%%%%%%%%%%%%%%%%%%%%%%%%%%%%%%%%%%%%%%%%%%%%%%
\begin{abstract} 

Survey responses on socially undesirable behaviors, such as self-reported spousal violence, are often subject to underreporting due to social stigma, fear of retaliation, and other reporting pressures. When such data are analyzed using standard econometric models that focus on  conditional means, such as probit and logit models, the resulting estimates are likely to be biased and can obscure heterogeneity in covariate effects. To address these challenges, we propose a Bayesian binary quantile regression framework that accounts for misclassification and provides quantile-specific effects for the latent true response. The framework incorporates false-negative and false-positive probabilities to capture reporting errors and employs a novel partially collapsed Gibbs sampler for estimation. We also discuss the computation of covariate effects and marginal likelihood for Bayesian model comparison. Simulation studies under various settings (prior effective sample size, misclassification rates, and prior distribution) show that accounting for misclassification improves inference across quantiles relative to models that ignore reporting errors. We apply the framework to women's self-reported spousal violence and find that underreporting of spousal violence exceeds overreporting across quantiles, while model comparisons using marginal likelihood generally favor the quantile model with misclassification and yield different conclusions about the determinants of reported violence.
\end{abstract}

{\texttt{Key words}: Quantile Regression, Markov chain Monte Carlo, Misclassification.}
\bigskip

%------------------------------------------------------------------------------
\section{Introduction}
%------------------------------------------------------------------------------

Spousal violence against women is a pervasive problem, but understanding its determinants and designing effective interventions remain challenging because of inconsistencies in how violence is measured and the limited availability of reliable data \citep{Clark-et-al-2023, Crawford-LloydLaney-2024}. Official statistics derived from law enforcement, health systems, and judicial records typically capture only a subset of cases that are formally reported or involve severe harm \citep{RuizPerez-Plazaola-VivesCases-2007}. Hence, population-based surveys are generally regarded as a more comprehensive source for estimating the prevalence and severity of spousal violence. However, survey data are not immune to measurement and reporting errors, as respondents may be affected by recall bias and systematic misreporting \citep{Bound-Brown-Mathiowetz-2001, Innes-et-al-2022}. These concerns are particularly pronounced in the context of spousal violence, given its sensitive and stigmatized nature, which can discourage disclosure and result in high non-response rates and substantial underreporting \citep{RuizPerez-Plazaola-VivesCases-2007, Cullen-2023, Polettini-etal-2024, Arima-etal-2026}. Despite the importance of underreporting in this context, the empirical literature on spousal violence has largely overlooked the resulting misclassification in reported outcomes. Among the limited studies that account for misclassification, the analysis has been confined to conditional-mean models, including Logit \citep{Chin-etal-2017}, Probit \citep{Wang-etal-2026}, and Poisson regression models \citep{Polettini-etal-2024, Arima-etal-2026}.

Quantile regression \citep{Koenker-Basset-1978}, unlike conditional mean models such as linear regression, models the conditional quantiles of the response variable as a function of covariates and offers a richer characterization of covariate effects across the outcome distribution. Extending quantile regression to discrete outcomes, however, presents both conceptual and methodological challenges. In the case of binary outcomes, the conditional distribution is discrete, making the interpretation and estimation of conditional quantiles nontrivial. Several approaches have been proposed to address this issue, including semiparametric binary median regression \citep{manski1975maximum,manski1985semiparametric} and smoothed binary quantile regression \citep{Kordas-2006}. Within the Bayesian approach, 
\citet{Benoit-Poel-2012} proposed a binary quantile regression framework and relied on the Metropolis–Hastings (MH) algorithm for its estimation. Whereas, \citet{Ji-etal-2012} developed binary quantile regression with variable selection and proposed a Gibbs sampling algorithm that utilized the normal–exponential mixture formulation of the asymmetric Laplace (AL) distribution. \citet{Ojha-Rahman-2021} adopts a similar framework and presents a Gibbs estimation algorithm without variable selection.

%Quantile regression \citep{Koenker-Basset-1978}, unlike conditional mean models such as linear regression, models the conditional quantiles of the response variable as a function of the covariates. This framework has been extensively used in economics, finance, and the social sciences, particularly to analyze the tails of the conditional distribution. Extending this approach to discrete outcomes results in conceptual and methodological challenges. For binary data, the conditional distribution is discrete, rendering the interpretation of conditional quantiles nontrivial. To address these challenges, several approaches have been proposed for applying quantile regression to binary response. \cite{manski1975maximum,manski1985semiparametric} developed a general semiparametric binary median (50th quantile) regression estimator. \cite{Kordas-2006} introduced a smoothed binary quantile regression approach that enables the estimation of covariate effects varying across the latent response distribution. \citet{Benoit-Poel-2012} proposed a Bayesian quantile regression framework for binary responses, but relied on the Metropolis–Hastings (MH) algorithm for estimation. Whereas, \citet{Ji-etal-2012} developed a binary quantile regression with variable selection that leverages the normal–exponential mixture representation of an asymmetric Laplace (AL) distribution to propose a Gibbs sampling algorithm. \citet{Ojha-Rahman-2021} adopts a similar framework and presents a Gibbs estimation algorithm without variable selection.

While the above-mentioned studies focus on binary quantile regression, they do not address the issue of misclassification in the binary response. Misclassification, however, is a pervasive concern in binary outcome data across a range of fields, including public health, economics, and quality control. This issue is particularly relevant to our application, where women’s self-reported experiences of spousal violence may be underreported and, consequently, misclassified because of social stigma, fear of repercussions, and other reporting pressures. Frequentist approaches to response misclassification typically assume that the misclassification parameters are known and treated as fixed, or that validation data are available for a subsample in which the true outcome is established using a more accurate classification procedure; see, for example, \citet{lyles2011validation}. Bayesian approaches can similarly incorporate validation data or prior information. For example, \citet{mcglothlin2008binary} account for response misclassification and covariate measurement error by combining validation data with weakly informative priors on the misclassification parameters. When validation data are unavailable, we instead follow \citet{McInturff-etal-2004} and incorporate expert information on misclassificaton parameters through informative priors. Centering these priors at values that would otherwise be treated as fixed allows us to incorporate uncertainty about the misclassification parameters directly into posterior inference.

The above discussion highlights that binary quantile regression and response misclassification has attracted considerable attention, but these two issues have not yet been considered jointly. To address this gap, we develop a Bayesian quantile regression framework for binary responses subject to misclassification. Rather than treating the false-negative and false-positive probabilities as known and fixed, we model them as unknown parameters, allowing uncertainty about the misclassification process to propagate through posterior inference. We introduce latent variables to facilitate data augmentation and develop a novel partially collapsed Gibbs (PCG) sampler \citep{Dyk-Park-2008} for efficient posterior computation. The proposed framework enables joint inference on the regression coefficients and misclassification probabilities while capturing heterogeneity in covariate effects across conditional quantiles. Finally, following \citet{Chib-1995}, we use the Gibbs output to compute the marginal likelihood for Bayesian model comparison and thus contribute to the growing literature on comparison of quantile regression model using marginal likelihood.

Our proposed binary quantile regression model with misclassification (BQRM) is evaluated through extensive simulation studies covering a range of settings, including different prior effective sample sizes, degrees of misclassification, and prior specifications. The results show that, relative to binary quantile regression (BQR) models that ignore misclassification, the BQRM yields lower bias and mean squared error (MSE), along with higher coverage rates. Marginal-likelihood-based model comparison also favors the BQRM over the BQR across quantiles. Overall, the BQRM outperforms the BQR model which ignores misclassification.

We then apply the proposed framework to data from India to examine the relationship between women’s self-reported spousal violence (widely known for underreporting) and female employment status, household wealth, and other socio-demographic characteristics. Three main findings emerge. First, the probability of a false negative is consistently higher than that of a false positive across quantiles, consistent with the spousal violence literature that underreporting is more prevalent than overreporting. Second, marginal likelihoods favor the BQRM over the BQR for most quantiles, providing further evidence that accounting for misclassification improves model fit. Third, accounting for misclassification changes the conclusions regarding the determinants of reported spousal violence. For instance, while female employment is positively associated with reporting spousal violence when misclassification is ignored, this association loses statistical credibility after accounting for misclassification, particularly at lower quantiles. In contrast, household wealth exhibits a consistently strong negative association with women’s reporting of spousal violence across the conditional quantiles.

The remainder of the paper is organized as follows. Section~\ref{sec:Data} presents the data for the study, while Section~\ref{sec:Method} describes the binary quantile model without and with misclassification along with their estimation procedure, computation of covariate effects and the marginal likelihood. Section~\ref{sec:Simulations} provides multiple simulation studies to examine the performance of the proposed approach relative to a BQR model that ignores misclassification. Section~\ref{sec:Application} implements the proposed method to study spousal violence as reported by women in India, and lastly, Section~\ref{sec:Conclusion} concludes with a discussion of the findings and directions for future research.

%------------------------------------------------------------------------------
\section{Data}\label{sec:Data}

%------------------------------------------------------------------------------

The 2005--06 National Family Health Survey (NFHS-3), conducted as part of the India Demographic and Health Survey (DHS) program, included the standardized Domestic Violence Module developed by DHS to measure the prevalence and severity of intimate partner violence in the population. The survey sought to minimize reporting errors by using behaviorally specific questions (e.g., whether the respondent had been slapped or punched) rather than abstract or judgmental questions (e.g., ``Has he ever physically harmed you?''). Nevertheless, self-reported measures of intimate partner violence remain susceptible to misclassification arising from underreporting, recall bias, social desirability bias, and respondents' interpretation of the questions. These features make the NFHS-3 data particularly well suited for applying our proposed Bayesian binary quantile regression framework with misclassification.

This study analyzes women's self-reported spousal violence in the previous 12 months as a function of several demographic and socioeconomic variables collected in the survey. The analysis focuses on currently married urban women who were administered the domestic violence module. After excluding observations with missing data, the final sample consists of 20,115 respondents. The primary outcome (vio) is an indicator of whether a woman experienced any of the following acts perpetrated by her husband during the past year: (1) slapping; (2) twisting her arm or pulling her hair; (3) pushing, shaking, or throwing an object at her; (4) punching with a fist or another harmful object; (5) kicking or dragging; (6) attempted choking or burning; or (7) threats or attacks with a knife, gun, or other weapon.

Explanatory variables include women's age (fage), women's employment status (fwork), husbands' years of education (meduc), the number of children in the household (nchildren), the number of other adult women in the household, indicators for women who have been married more than once (remarriage) and for women in polygynous unions (polyg), the number of other adult women in the household (nwomen), and the household wealth index (wealth). The DHS wealth index is a composite measure of long-term household socioeconomic status constructed from information on housing characteristics, utilities, and durable assets. Using principal components analysis, DHS generates a weighted score that ranks households on a relative socioeconomic scale.

Table \ref{india_summary} presents descriptive statistics for the variables used in the analysis. On average, 17 percent of women report experiencing physical violence by their husband in the past year. The mean age of women in the sample is 32, and approximately 28 percent worked outside the household in the previous year. Roughly 2 percent of women report having been married more than once, and about 1 percent are in polygynous marriages. The average level of male education is about 9 years. Households contain, on average, two to three children and one to two other adult women. In our analysis, all continuous covariates were standardized to improve numerical stability and convergence.

%------------------------------------------------------------------------
\begin{table}[!ht]
\centering
\caption{Descriptive summary. Standard deviation (Std) reported only for non-indicator variables.}
\label{india_summary}
\begin{tabular}{lp{9cm}rr}
\toprule
Variable & Description &Mean & Std\\ 
\midrule
vio   & dependent variable, indicator for spousal violence &0.17 &  \\ 
fage  & women's age (in years) &32.38 & 7.71 \\ 
fwork & indicator variable for a woman's employment status outside the household &0.28 &  \\ 
meduc & husband's education (in years) & 9.23 & 5.10 \\ 
wealth & household wealth index & 0.71 & 0.83 \\ 
nchildren & number of children in the household & 2.20 & 1.40 \\ 
remarriage & indicator variable for women who have been married more than once &0.02 &  \\ 
polyg &  indicator variable for women in polygynous unions & 0.01 &  \\ 
nwomen & number of other adult women in the household &1.59 & 0.86 \\ 
\bottomrule
\end{tabular}
\end{table}
%------------------------------------------------------------------------

%------------------------------------------------------------------------------
\section{Methodology}\label{sec:Method}
%------------------------------------------------------------------------------

This section develops the binary quantile regression model under both correct classification and misclassification and outlines their estimation procedures using Markov chain Monte Carlo (MCMC) methods. We first consider the standard model with correctly observed binary outcomes and then extend it to allow reporting errors that cause observed outcomes to differ from their true underlying values. in which binary outcomes are observed without error. We conclude by describing the computation of covariate effects and marginal likelihoods for Bayesian model comparison.

%------------------------------------------------------------------------------
\subsection{Binary Quantile Regression}\label{sec:Model1}
%------------------------------------------------------------------------------

Quantile regression \citep{Koenker-Basset-1978}, or its Bayesian 
counterpart \citep{Yu-Moyeed-2001}, models the conditional quantiles of a 
continuous response variable as a function of covariates (explanatory or 
independent variables). While the former does not make any distributional 
assumption on the error term, the latter creates a working likelihood 
assuming the errors follow an asymmetric Laplace (AL) distribution 
\citep{Yu-Zhang-2005}. However, when the response variable is discrete—such 
as binary, ordinal, or count—the estimation of quantiles becomes 
problematic, as the quantile function is not well-defined due to the
degeneracy of the distribution. A widely adopted solution is to leverage the 
equivariance property of quantiles\footnote{Monotonic equivariance in 
quantile regression refers to the property that quantiles are preserved under 
monotonic (i.e., strictly increasing or decreasing) transformations of the 
response variable. For example, let $Q_{y}(p|x) = \beta_{1} + \beta_{2} x$ be 
the conditional quantile function of a continuous random variable $y$ at 
quantile $p$. Consider the transformation $z=\ln(y)$, which is a 
monotonically increasing function. By the monotonic equivariance property, 
the conditional quantile function of $z$ becomes $Q_{z}(p|x) = 
\ln(Q_{y}(p|x))=\ln(\beta_{1} + \beta_{2} x)$. Although 
$Q_{z}(p|x)$ is no longer linear in $x$, the quantile structure is preserved 
through the transformation. This property plays a crucial role in Bayesian 
quantile regression as the \textit{cdf} 
of the AL distribution is monotonic, supporting quantile inference under 
monotonic transformations.} and employ the
latent variable framework \citep{Albert-Chib-1993}, wherein the observed 
discrete outcome is assumed to arise from an unobserved continuous latent 
process, and inference is performed on the quantiles of this latent 
construct. This framework has been extensively applied in Bayesian quantile 
regression across various contexts, including binary outcomes in both 
cross-sectional \citep{Kordas-2006, Benoit-Poel-2012} and panel data models 
\citep{Rahman-Vossmeyer-2019, Bresson-etal-2021}, as well as ordinal outcomes 
in cross-sectional \citep{Rahman-2016, Maheshwari-Rahman-2023} and 
longitudinal settings \citep{Alhamzawi-Ali-Longitudinal2018}. The combination 
of monotonic equivariance and the latent variable approach provides a 
coherent and principled basis for estimation and inference in quantile 
regression models involving discrete outcomes.

To illustrate this framework more concretely, consider the case of binary 
response data. The binary quantile regression (BQR) model can be conveniently 
represented in terms of the latent variable $z$ as follows,
%-------------------------------
\begin{equation}
\begin{split}
z_{i} & =  x'_{i} \beta_{p}  + \varepsilon_{i}, \hspace{0.75in} \forall \;
i=1, \cdots, n, \\
y_{i} & = \left\{ \begin{array}{ll}
1 & \textrm{if} \;  z_{i} > 0,\\
0 & \textrm{otherwise}.
\end{array} \right.
\end{split}
\label{eq:Model1}
\end{equation}
%-------------------------------
where $z_{i}$ is the latent variable corresponding to unit (individual, 
household, firm) $i$, $x_{i}$ is a $k \times 1$ vector of covariates, 
$\beta_{p}$ is a $k \times 1$ vector of unknown parameters at the $p$-th 
quantile (henceforth, the subscript $p$ is dropped for notational 
convenience), $\varepsilon_{i}$ follows an AL distribution i.e., $\varepsilon_{i} 
\sim AL(0,1,p)$, and $n$ denotes the number of observations\footnote{Binary response models typically require both location and scale restrictions for parameter identification \citep{Jeliazkov-Rahman-2013}. As in the Probit and Logit models, the location restriction in binary quantile regression is imposed by fixing the (first and only) cut-point at zero. The scale restriction is achieved by setting the scale parameter $\sigma=1$, which implies the variance is constant for any quantile $p$. For example, at the median i.e., $p=0.5$, we have $V(\varepsilon)=8$. This differs from the Probit and Logit models, where the latent error variances are fixed at $1$ and $\pi^{2}/3$, respectively. The same identification scheme applies to the binary quantile regression with misclassification discussed in the next section.}. The latent 
variable $z$ can be interpreted in terms of latent utility differential or 
some kind of propensity. In our application, it can be interpreted as a woman's 
propensity to report violence committed by her spouse. When the observed (or reported) response $y_{i}=1$ ($y_{i}=0$), propensity to report spousal violence is likely to be high (low) and $z_{i}$ takes a value in the positive (negative) part of the real line. In the absence of misreporting, the propensity to report spousal violence coincides with the underlying propensity of the spouse to commit violence against the woman.

For our binary quantile model defined in~\eqref{eq:Model1}, it can be shown 
that $\Pr(y_{i}=1) = \Pr(z_{i} > 0) = 1 - F_{AL}(-x'_{i}\beta,1,p)$, where 
$F_{AL}(\cdot)$ denotes the cumulative 
distribution function (\textit{cdf}) of the AL distribution with location 
$-x'_{i}\beta$, scale $\sigma=1$, and quantile $p$. This relationship allows 
for the construction of a working likelihood and enables estimation of the 
binary quantile model using an MCMC algorithm  
\citep{Benoit-Poel-2012}, but it is not amenable to constructing a pure Gibbs 
sampler \citep{Geman-Geman-1984}. A more efficient 
approach, following \citet{Kozumi-Kobayashi-2011}, is to express the error 
term in normal-exponential mixture formulation: $\varepsilon_{i} = 
\theta w_{i} + \tau \sqrt{w_{i}} \,u_{i}$, and rewrite the binary quantile 
model as, 
%-------------------------------
\begin{equation}
\begin{split}
z_{i} & =  x'_{i} \beta  + \theta w_{i} + \tau \sqrt{w_{i}} \,u_{i},
\hspace{0.75in} \forall \; i=1, \cdots, n, \\
y_{i} & = \left\{ \begin{array}{ll}
1 & \textrm{if} \;  z_{i} > 0,\\
0 & \textrm{otherwise}
\end{array} \right.
\end{split}
\label{eq:Model2}
\end{equation}
%-------------------------------
where $\theta = \frac{(1-2p)}{p(1-p)}$, $\tau = \sqrt{\frac{2}{ p(1-p)}}$, 
and $w_{i} \sim \mathcal{E}(1)$ is independently distributed
of $u_{i} \sim N(0,1)$. Here, the notations $\mathcal{E}$ and $N$ denote
exponential and normal distributions, respectively. It is clear from
formulation \eqref{eq:Model2} that the latent variable $z_{i}|\beta,w_{i}
\sim N( x'_{i}\beta + \theta w_{i}, \tau^{2} w_{i})$ which enables us to leverage the properties of normal distribution and construct a pure Gibbs algorithm.

%-------------------------------------------------------------------------------
\begin{table*}[!t]
\begin{algorithm}[Gibbs Algorithm for Binary Quantile Regression (BQR)]
\label{alg:algorithm1}
\rule{\textwidth}{0.5pt} \normalsize{
\begin{enumerate}
\item    Sample $\beta| z,w$ $\sim$  $N(\tilde{\beta}, \tilde{B})$, where,
\item[]  $\tilde{B}^{-1} = \bigg(\sum_{i=1}^{n}
        \frac{x_{i} x'_{i}}{\tau^{2} w_{i}} + B_{0}^{-1} \bigg) $ \hspace{0.05in}
         and \hspace{0.05in} $\tilde{\beta} = \tilde{B}\bigg( \sum_{i=1}^{n}
         \frac{x_{i}(z_{i} - \theta w_{i})}{\tau^{2} w_{i}} + B_{0}^{-1} \beta_{0}
         \bigg)$.
%------------------------------------------------------------------------------
\item    Sample $w_{i}|\beta, z_{i}$ $\sim$  $GIG \, (0.5, \tilde{\lambda}_{i},
    \tilde{\eta}) $, for $i=1,\cdots,n$, where,
\item[]  $\tilde{\lambda}_{i} = \Big( \frac{ z_{i} - x'_{i}\beta}{\tau}
    \Big)^{2}$ \hspace{0.05in} and \hspace{0.05in} $\tilde{\eta} = \Big(
    \frac{\theta^{2}}{\tau^{2}} + 2 \Big)$.
%------------------------------------------------------------------------------
\item    Sample the latent variable $z_{i}|y_{i},\beta,w_{i}$ for
         all values of $i=1,\cdots,n$, from an univariate
         truncated normal (TN) distribution as follows,
         \begin{eqnarray*}
         z_{i}|y_{i}, \beta,w_{i} & \sim & \left\{
         \begin{array}{ll}
         TN_{(-\infty,0]}\Big(x'_{i}\beta + \theta w_{i},
         \tau^{2} w_{i} \Big)
         & \textrm{if} \;\;  y_{i} = 0,\\ [0.9em]
         TN_{(0,\infty)}\Big(x'_{i}\beta + \theta w_{i},
         \tau^{2} w_{i} \Big)
         & \textrm{if} \;\; y_{i} = 1.
         \end{array}
         \right.
         \end{eqnarray*}
\end{enumerate}}
\rule{\textwidth}{0.5pt}
\end{algorithm}
\end{table*}
%-------------------------------------------------------------------------------

By Bayes' theorem, the complete data likelihood~\eqref{eq:Model2} is combined with a normal prior: $\beta \sim N(\beta_{0}, B_{0})$, to form the complete data posterior distribution \citep{Tanner-Wong-1987}. This yields the following expression,
%-------------------------------
\begin{equation}
\begin{split}
\pi(z,\beta,w|y)
          & \propto  \bigg\{ \prod_{i=1}^{n}
          \big[ I(z_{i}>0)I(y_{i}=1) + I(z_{i} \leq 0) I(y_{i}=0) \big]
          \; N(z_{i}|x'_{i}\beta + \theta w_{i}, \tau^{2} w_{i}) \\
          & \times \; \mathcal{E}(w_{i}|1) \bigg\}
          \; N(\beta_{0}, B_{0}).
          \label{eq:CompDataPost}
\end{split}
\end{equation}
%-------------------------------
The full conditional posterior densities for $(z, \beta, w)$ can be derived 
from equation~\eqref{eq:CompDataPost}, and the model can be estimated using 
the Gibbs sampler outlined in Algorithm~\ref{alg:algorithm1}. This algorithm 
was originally presented in \citet{Ojha-Rahman-2021}, and can be considered 
as a simplified version (without variable selection) of \citet{Ji-etal-2012}. 
Estimation proceeds by iteratively sampling the parameters from their full 
conditional distributions. Specifically, the regression coefficients $\beta$ are sampled from a multivariate normal distribution, the latent weights $w$ are sampled element-wise from a Generalized Inverse Gaussian (GIG) distribution \citep{Hormann-Leydold-2014}, and the latent variables $z$ are sampled from a truncated normal distribution \citep{Robert-1995}.

%-------------------------------------------------------------------------------
\subsection{Binary Quantile Regression with 
Misclassification}\label{sec:Model2}

When misclassification is present in the data, the true binary response 
variable $y \in \{0, 1\}$ as introduced in equation~\eqref{eq:Model1} is not 
observed. Instead, we observe a misclassified version $y^{obs}$, in 
which some actual 1's are incorrectly reported (and hence recorded) as 0's, and some true 0's are incorrectly recorded as 1's. The probability of the former$-$known as false negative$-$is denoted by $\delta_{01} = \Pr(y^{obs}=0|y=1)$, while the probability of the latter$-$referred to as false positive$-$is given by 
$\delta_{10} = \Pr(y^{obs}=1|y=0)$. Two associated measures are 
\emph{sensitivity}, defined as $\Pr(y^{obs}=1|y=1) = 1-\delta_{01}$, and 
\emph{specificity} as $\Pr(y^{obs}=0|y=0) = 1-\delta_{10}$, which represents 
the probabilities of correctly identifying `true positives' and `true 
negatives', respectively. So, the relationship between $y^{obs}$ and $y$ can 
be expressed through the equation, 
%-----------------------
\begin{displaymath}
y^{obs}_{i} = y_{i} + \nu_{i}, \quad \mathrm{where} \quad 
	\nu_{i} = \left\{ \begin{array}{ll}
		-1 & \textrm{with probability $\delta_{01}$ if $y_{i}=1$},\\
		1 & \textrm{with probability $\delta_{10}$ if $y_{i}=0$},\\
		0 & \textrm {otherwise}.
	\end{array} \right.
\end{displaymath}
%-----------------------
Naturally, the presence of misclassification requires re-specifying the likelihood, joint posterior distribution, and formulating a suitable MCMC algorithm for model estimation.

For the binary quantile regression with misclassification (BQRM), the full 
likelihood $L(\Theta|y^{obs})$ can be shown to have the following 
expression,
%--------------------------
\begin{equation}
\begin{split}
L( \Theta|y^{obs}) & = \prod_{i=1}^{n} 
\bigg\{ 
\Pr(y^{obs}_{i}=1)^{y_{i}^{obs}} \Big[ 1 - \Pr(y^{obs}_{i}=1)  
\Big]^{(1- {y_{i}^{obs})}} \bigg\} \\
& =  \prod_{i=1}^{n} \bigg\{ \Big[ \delta_{10} F_{AL}(-x'_{i}\beta) + 
(1-\delta_{01}) (1 - F_{AL}(-x'_{i}\beta))  \Big]^{y_{i}^{obs}} 
\\
& \qquad \times \Big[ (1-\delta_{10}) F_{AL}(-x'_{i}\beta) + \delta_{01} (1 - 
F_{AL}(-x'_{i}\beta))  \Big]^{(1- {y_{i}^{obs})}}   \bigg\},
\end{split}
\label{eq:bqrMis-fullLike} 
\end{equation}
%--------------------------
where $\Theta=(\beta,\delta_{10},\delta_{01})$, $\Pr(y_{i}=1) = 
1-F_{AL}(-x'_{i}\beta)$, and recall $F_{AL}(\cdot)$ is the \textit{cdf} of a standard AL distribution. However, the full likelihood~\eqref{eq:bqrMis-fullLike} is not conducive to constructing an MCMC algorithm, as observed in 
Section~\ref{sec:Model1}.

To facilitate a convenient formulation inspired from \citep{Tu-etal-1999}, we 
utilize both the latent variables $z$ and $y$$-$with the latter now 
considered latent since its unobserved$-$and write the complete data likelihood as,
%--------------------------
\begin{equation}
\begin{split}
	&L(\Theta,z,y,w|y^{obs})  
	%\prod_{i=1}^{n} \bigg\{ 
	%f(y_{i}^{obs}|y_{i},\delta_{01},\delta_{10}) f(y_{i}|z_{i})  
	%f_{AL}(z_{i}|\beta)\bigg\} \\
	= \prod_{i=1}^{n} \bigg\{ 
	f(y_{i}^{obs}|y_{i},\delta_{01},\delta_{10}) f(y_{i}|z_{i})  
	f_{N}(z_{i}|\beta,w_{i}) f_{\mathcal{E}}(w_{i})\bigg\} \\
	& = \prod_{i=1}^{n} \bigg\{    
	\big[ 	\delta_{01}^{y_{i}(1-y^{obs}_{i})} 
	(1-\delta_{01})^{y_{i}^{obs} y_{i}}   \times 
	\delta_{10}^{y^{obs}_{i}(1-y_{i})} 
	(1-\delta_{10})^{(1-y_{i}^{obs})(1-y_{i})} \big] \\
	& \quad \times \big[ I(z_{i}>0)I(y_{i}=1) + I(z_{i} \leq 
	0) I(y_{i}=0) \big] %f_{AL}(z_{i}|\beta).
    f_{N}(z_{i}|\beta,w_{i}) f_{\mathcal{E}}(w_{i}) \bigg\},
\end{split}
\label{eq:bqrMis-AugLike} 
\end{equation}
%--------------------------
where $f_{N}(\cdot)$ and $f_{\mathcal{E}}(\cdot)$ denote the densities of 
normal and exponential distributions, respectively. In 
equation~\eqref{eq:bqrMis-AugLike}, 
$f(y_{i}^{obs}|y_{i},\delta_{01},\delta_{10})$ is the misclassification 
component that corresponds to the expression within first square bracket 
(second line), and $f(y_{i}|z_{i})$ is simply an indicator function 
independent of any parameter that appears in the second square bracket. 
Furthermore, equation~\eqref{eq:bqrMis-AugLike} utilizes mixture 
representation of the AL density i.e., $f_{AL}(z|\beta) = \int 
f_{N}(z|\beta,w) f_{\mathcal{E}}(w) \, dw$ \citep{Kotz-etal-2001}. The 
mixture representation, also used
in binary quantile regression (without misclassification) in 
Section~\ref{sec:Model1}, proves useful for 
estimating binary quantile models with misclassification.

%-------------------------------------------------------------------------------
\begin{table*}[!t]
\begin{algorithm}[Partially Collapsed Gibbs Algorithm for BQRM] 
\label{alg:algorithm2}
\rule{\textwidth}{0.5pt} \normalsize{
\begin{enumerate}
\item    Sample $\beta| z,w$ $\sim$  $N(\tilde{\beta}, \tilde{B})$, where,
\item[]  $\tilde{B}^{-1} = \bigg(\sum_{i=1}^{n}
\frac{x_{i} x'_{i}}{\tau^{2} w_{i}} + B_{0}^{-1} \bigg) $ \hspace{0.05in}
and \hspace{0.05in} $\tilde{\beta} = \tilde{B}\bigg( \sum_{i=1}^{n}
\frac{x_{i}(z_{i} - \theta w_{i})}{\tau^{2} w_{i}} + B_{0}^{-1} \beta_{0}
\bigg)$.

%------------------------------------------------------------------------------
\item    Sample $\delta_{01}|y,y^{obs}$ $\sim$  
$Beta (\tilde{\kappa}_{1}, \tilde{\kappa}_{2})$ 
where,
\item[]  $\tilde{\kappa}_{1} = \Big( \sum_{i=1}^{n} y_{i}(1 - y^{obs}_{i}) + 
\kappa_{1} \Big)$ \hspace{0.05in} and \hspace{0.05in} $\tilde{\kappa}_{2} = 
\Big( \sum_{i=1}^{n} y_{i} y^{obs}_{i} + \kappa_{2} \Big)$.
%------------------------------------------------------------------------------
\item    Sample $\delta_{10}|y,y^{obs}$ $\sim$  
$Beta (\tilde{\kappa}_{3}, \tilde{\kappa}_{4})$ 
where,
\item[]  $\tilde{\kappa}_{3} = \Big( \sum_{i=1}^{n} y^{obs}_{i}(1 - y_{i}) + 
\kappa_{3} \Big)$ \hspace{0.05in} and \hspace{0.05in} $\tilde{\kappa}_{4} = 
\Big( \sum_{i=1}^{n} (1 - y_{i}) (1 - y^{obs}_{i}) + \kappa_{4} \Big)$.
%------------------------------------------------------------------------------
\item  We marginalize the conditional posterior distribution of $y$ over 
$(z,w)$ and sample each $y_{i}|\beta,\delta_{01},\delta_{10}, y^{obs}_{i}$ 
$\sim$  $Bernoulli (\Psi_{i})$ for all values of $i=1, \cdots,n$, where,
\item[] $\Psi_{i} =  \frac{\delta_{01}^{(1-y^{obs}_{i})} 
	(1-\delta_{01})^{y^{obs}_{i}} \times  
	\psi_{i}}{\delta_{01}^{(1-y^{obs}_{i})} 
	(1-\delta_{01})^{y^{obs}_{i}} \times  \psi_{i} \, + \, 
	\delta_{10}^{y^{obs}_{i}} 
	(1-\delta_{10})^{(1-y^{obs}_{i})} \times  (1-\psi_{i} ) } $, \quad
	and \quad $\psi_{i} = 1- F_{AL}(-x'_{i}\beta)$. 
%------------------------------------------------------------------------------
\item    Sample $w_{i}|\beta, z_{i}$ $\sim$  $GIG \, (0.5, 
\tilde{\lambda}_{i},
\tilde{\eta}) $, for $i=1,\cdots,n$, where,
\item[]  $\tilde{\lambda}_{i} = \Big( \frac{ z_{i} - x'_{i}\beta}{\tau}
\Big)^{2}$ \hspace{0.05in} and \hspace{0.05in} $\tilde{\eta} = \Big(
\frac{\theta^{2}}{\tau^{2}} + 2 \Big)$.
%------------------------------------------------------------------------------
\item    Sample $z_{i}|\beta,w_{i},y_{i}$ for
all values of $i=1,\cdots,n$, from a univariate
truncated normal (TN) distribution as follows,
\begin{eqnarray*}
	z_{i}|\beta,w_{i},y_{i} & \sim & \left\{
	\begin{array}{ll}
		TN_{(-\infty,0]}\Big(x'_{i}\beta + \theta w_{i},
		\tau^{2} w_{i} \Big)
		& \textrm{if} \;\;  y_{i} = 0,\\ [0.9em]
		TN_{(0,\infty)}\Big(x'_{i}\beta + \theta w_{i},
		\tau^{2} w_{i} \Big)
		& \textrm{if} \;\; y_{i} = 1.
	\end{array}
	\right.
\end{eqnarray*}
\end{enumerate}}
\rule{\textwidth}{0.5pt}
\end{algorithm}
\end{table*}
%-------------------------------------------------------------------------------

We complete the Bayesian setup by assigning prior distributions to the parameters: $\beta \sim N(\beta_{0}, B_{0})$, $\delta_{01} \sim 
B(\kappa_{1}, \kappa_{2})$, and $\delta_{10} \sim B(\kappa_{3},\kappa_{4})$, 
where $B(\cdot)$ denotes a Beta distribution. Applying Bayes' theorem, we 
combine the complete data likelihood~\eqref{eq:bqrMis-AugLike} with the prior 
distributions on $(\beta, \delta_{01}, \delta_{10})$ to obtain the
joint posterior distribution, which, as usual, is analytically 
intractable. Consequently, we derive the conditional posteriors and propose a 
novel partially collapsed Gibbs (PCG) sampling algorithm to estimate the misclassified binary quantile model. While the detailed derivations are provided in the \ref{app:full}, the PCG sampling scheme is outlined in Algorithm~\ref{alg:algorithm2}.
 
The proposed algorithm updates each parameter and latent variable iteratively from its full conditional distribution, with judicious marginalization to improve mixing and accelerate convergence. The regression coefficients $\beta$ are drawn from a multivariate normal distribution, while the misclassification parameters $\delta_{01}$ and $\delta_{10}$ are updated from their respective Beta conditionals. The key step in the partially collapsed construction is the update of the true binary response $y$.  Rather than conditioning on all latent variables, we marginalize the conditional density over $(z, w)$ and sample each $y_i$ from a Bernoulli distribution. This targeted marginalization partially collapses the Gibbs sampler, reducing posterior dependence among the sampled quantities and thereby improving mixing and convergence. Conditional on $y$, the latent weights $w$ and the continuous latent variable $z$ are then sampled element-wise from truncated normal and GIG distributions, respectively. 

%Together, these steps constitute a PCG sampling scheme that enables efficient exploration of the posterior distribution despite the model's inherent complexity.

%-------------------------------------------------------------------------------
%------------------------------------------------------------------------------
\subsection{Covariate Effects}\label{sec:CovEff}
%------------------------------------------------------------------------------

Both binary quantile models, with and without misclassification, are nonlinear in the parameters, so the regression coefficients do not directly represent covariate effects. We therefore consider marginal effects on the probability of success. For the $l$-th covariate, let $\{x_{i,l}\}$ take two values $a$ 
and $b$, denoted as $\{x_{i,l}^{a}\}$ and $ \{ x_{i,l}^{b}\}$, respectively. 
We partition the covariate and parameter vectors as: $x_{i}^{a} = ( 
x_{i,l}^{a}, \, x_{i,-l})$, $x_{i}^{b} = (x_{i,l}^{b}, \, x_{i,-l})$, and 
$\beta_{p} = (\beta_{p,l}, \, \beta_{p,-l})$, where $-l$ in the subscript 
denotes all covariates (parameters) except the $l$-th covariate (parameter). 
Once again, we suppress the subscript $p$ (denoting the quantile) to 
maintain consistency with earlier sections. We first describe the computation of average covariate effect for the model without misclassification and then extend it to the misclassified model.

For the covariate effect in binary quantile model without misclassification, 
we are interested in the distribution of $\{\Pr(y_{i}=1|x_{i,l}^{b}) - 
\Pr(y_{i}=1|x_{i,l}^{a} ) \}$. Note that this quantity is marginalized 
over the remaining covariates $\{x_{i,-l}\}$ and the parameter $\beta$, 
but is conditional on the data $y=(y_{1}, \cdots, y_{n})'$. We marginalize 
the covariates using their empirical distribution and the parameters based on 
the posterior distribution.

To generate draws from the distribution of $\{\Pr(y_{i}=1|x_{i,l}^{b}) -
\Pr(y_{i}=1|x_{i,l}^{a} ) \}$, we use the method of composition
\citep{Chib-Jeliazkov-2006,Rahman-Vossmeyer-2019, Bresson-etal-2021}. 
Specifically, for each iteration, we randomly select an individual, retain 
the observed values of the remaining covariates, draw a parameter vector 
$\beta$ from its posterior distribution, and evaluate
$\{\Pr(y_{i}=1|x_{i}^{b},\beta) -
\Pr(y_{i}=1|x_{i}^{a},\beta)\}$, where $ \Pr(y_{i}=1|x_{i}^{q},\beta) 
= \big[1 - F_{AL}(- x_{i,l}^{q} \, \beta_{l} - x'_{i,-l} \, \beta_{-l}) 
\big]$, for $q=b,a$. Repeating this procedure for all individuals and 
posterior draws yields the average covariate effect,
%-------------------------------------
\begin{equation}\label{eq:ACE1}
\frac{1}{G} \frac{1}{n} \sum_{g=1}^{G}
\sum_{i=1}^{n}
\Big[ \Pr(y_{i}=1|x_{i}^{b},\beta^{(g)}) -
\Pr(y_{i}=1|x_{i}^{a},\beta^{(g)})
\Big],
\end{equation}
%----------------------------------
where $\beta^{(g)}$ is a Gibbs draw of $\beta$ and 
$G$ is the number of post burn-in Gibbs draws.

%-------------------------------------
When evaluating the covariate effect in binary quantile model with 
misclassification, we have the parameters $\beta$ and $(\delta_{01}, \delta_{10})$ representing misclassification probabilities. Our distribution 
of interest, while intuitively unchanged, is now,
%-------------------------------------
\begin{equation*}
\{\Pr(y^{obs}_{i}=1|x_{i,l}^{b}) - 
\Pr(y^{obs}_{i}=1|x_{i,l}^{a} ) \},
\end{equation*}
%----------------------------------
and the marginalization is over $\{x_{i,-l}\}$ and 
$\Theta=(\beta, \delta_{01},\delta_{10})$. As before, 
covariates are integrated using their empirical distribution and parameters 
using their posterior distribution.

To generate draws from the distribution of 
$\{\Pr(y^{obs}_{i}=1|x_{i,l}^{b}) - \Pr(y^{obs}_{i}=1|x_{i,l}^{a} ) \}$, once 
again, we use the method of composition. This implies, for each retained draw 
$\Theta^{(g)}$ and across individuals, we compute 
$\{\Pr(y^{obs}_{i}=1|x_{i}^{b},\Theta) - 
\Pr(y^{obs}_{i}=1|x_{i}^{a},\Theta)\}$, 
where 
%-------------------------------------
\begin{equation*}
%\begin{split}
\Pr(y^{obs}_{i}=1|x_{i}^{q},\Theta) = \Big[\delta_{10} F_{AL}(- 
{x_{i}'}^{q} \,\beta) + (1 - \delta_{01}) 
\big( 1 - F_{AL}(- {x_{i}'}^{q}\,\beta) \big) \Big],
%\end{split}
\end{equation*}
%-------------------------------------
${x_{i}'}^{q} \,\beta = x_{i,l}^{q} \, \beta_{l} - x'_{i,-l} 
\, \beta_{-l}$, and $q=b,a$. Averaging the resulting probability 
differences across individuals and posterior draws gives the average 
covariate effect,
%-------------------------------------
\begin{equation}\label{eq:ACE2}
	\frac{1}{G} \frac{1}{n} \sum_{g=1}^{G}
	\sum_{i=1}^{n}
	\Big[ 
	\Pr(y^{obs}_{i}=1|x_{i}^{b},\Theta^{(g)})
	 -
	\Pr(y^{obs}_{i}=1|x_{i}^{a},\Theta^{(g)})
	\Big],
\end{equation}
%----------------------------------
where $\Theta^{(g)}$ is a Gibbs draw of $\Theta$ and $G$ is the number of post burn-in Gibbs draws.

%------------------------------------------------------------------------------
%------------------------------------------------------------------------------
\subsection{Marginal Likelihood}\label{sec:ML}

The Bayesian approach to model comparison is based on marginal likelihood 
\citep{Chib-1995} or equivalently 
through Bayes factor which is a ratio of marginal likelihoods. By averaging 
the likelihood function over parameter space, the marginal likelihood naturally 
embodies Occam’s razor, favoring parsimonious models over more complex models unless the latter provide a substantially better fit. Despite its central role in 
Bayesian inference, the use of marginal likelihood for comparing quantile 
regression models remains limited, notable exceptions include \citet{Yu-Stander-2007}, \citet{Kobayashi-Kozumi-2012} and \citet{Maheshwari-Rahman-2023}. 

When comparing binary quantile regression models with and without 
misclassification, it is important to note that such comparisons are valid 
only within a given quantile and not across quantiles. Moreover, if 
misclassification is present, the model without misclassification is 
inherently misspecified. A higher marginal likelihood for the 
model that accommodates misclassification therefore does not necessarily 
imply superior fit, but rather that the 
observed data are more plausibly generated by a latent-response process with 
measurement error than by a model assuming perfectly observed responses.

To compute the marginal likelihood for the binary quantile models, we follow 
the MCMC approach of \citet{Chib-1995}. The approach utilizes the identity that the marginal likelihood can be written as the ratio of the likelihood function times the prior density to the posterior density. So, the marginal likelihood $m(y|\mathcal{M})$ for model $\mathcal{M}$ is given by, 
%-----------------------------
\begin{equation}
	m(y|\mathcal{M}) = \frac{f(y|\mathcal{M},\Theta) \,
		\pi(\Theta|\mathcal{M})}{\pi(\Theta|\mathcal{M},y)}.
	\label{eq:BMLI}
\end{equation}
%-----------------------------
where $\Theta$ are parameters corresponding to model $\mathcal{M}$. 
The above equality holds for all values in the parameter space, but it 
typically computed at a high-density point (e.g., posterior mean or
mode) denoted $\Theta^{\ast}$ to minimize estimation variability. The
likelihood ordinate $f(y|\mathcal{M},\Theta^{\ast})$ is directly
available from the model, while the prior density
$\pi(\Theta^{\ast}|\mathcal{M})$ is specified by the researcher. Therefore, the main computational task is the estimation of the posterior ordinate 
$\pi(\Theta^{\ast}|y,\mathcal{M})$, which is obtained using the Gibbs 
outputs. Putting all components together, the log (logarithm) of marginal likelihood is estimated as,
%-----------------------------
\begin{equation}
	\ln \hat{m}(y) = \ln f(y|\Theta^{\ast}) + \ln \pi(\Theta^{\ast})
	- \ln \hat{\pi}(\Theta^{\ast}|y),
	\label{eq:logBMLI}
\end{equation}
%-----------------------------
where the conditioning on $\mathcal{M}$ is omitted for notational
simplicity. The next two subsections explain the computation of marginal
likelihood for binary quantile models without and with misclassification.

%------------------------------------------------------------------------------
\subsubsection{Marginal likelihood: Binary Quantile Regression}

The Gibbs sampling algorithm for estimating binary quantile regression (see 
Algorithm~1 in Section~\ref{sec:Model1}) is defined by the conditional 
posterior densities: $\pi(\beta|z,w)$, $\pi(w|\beta,z)$, and 
$\pi(z|\beta,w,y)$. Since, $(w,z)$ are latent variables, they are integrated 
out to reduce the dimension and we obtain the posterior ordinate as, 
%-----------------------------
\begin{equation*}
	\begin{split}
		\pi(\beta^{\ast}|y) & = \int 
		\pi(\beta^{\ast}|w,z,y) \,
		\pi(w,z|y) \, dw \, dz, 
	\end{split}
	\label{eq:POModel1}
\end{equation*}
%-----------------------------
The above can be estimated as the ergodic average of the conditional posterior density,
%-----------------------------
\begin{equation}
	\hat{\pi}(\beta^{\ast}|y) = G^{-1} \sum_{g=1}^{G}
	\pi(\beta^{\ast}|w^{(g)}, z^{(g)},y ).
	\label{eq:EstPOModel1}
\end{equation}
%-----------------------------
where $\beta^{\ast}$ denotes a high density point, such as the posterior mean 
or median. Substituting equation~\ref{eq:EstPOModel1} into 
equation~\eqref{eq:logBMLI}, gives an estimate of log marginal likelihood,
%-----------------------------
\begin{equation}
	\ln \hat{m}(y) = \ln f(y|\beta^{\ast})
	+ \ln \pi(\beta^{\ast}) 
	- \ln \hat{\pi}(\beta^{\ast}|y),
	\label{eq:EstMLModel1}
\end{equation}
%-----------------------------
where the full likelihood function $f(y|\beta^{\ast}) = \prod_{i=1}^{n}
\left[1 - F_{AL}(-x_i'\beta^{\ast})\right]^{y_i}
\left[F_{AL}(-x_i'\beta^{\ast})\right]^{1-y_i}$, and the prior normal density $\pi(\beta^{\ast})$ are evaluated at $\beta^{\ast}$.

%------------------------------------------------------------------------------
\subsubsection{Marginal likelihood: Binary Quantile Regression with 
Misclassification}

As shown in Section~\ref{sec:Model2}, the Gibbs sampling 
algorithm for BQRM involves six conditional posterior distributions. However, three of these correspond to the mixture or latent variables $(w,z,y)$. Once again, to reduce computational burden with high-dimensional integration, we integrate out these latent variables and express the posterior ordinate as,
%-----------------------------
\begin{equation}
	\begin{split}
		\pi(\beta^{\ast},\delta_{01}^{\ast},\delta_{10}^{\ast}|y^{obs})  =  
		\pi(\beta^{\ast}|y^{obs}) \, 
		\pi(\delta_{01}^{\ast}|y^{obs}) \, 
		\pi(\delta_{10}^{\ast}|y^{obs}).
	\end{split}
	\label{eq:POModel2}
\end{equation}
%-----------------------------
This factorization arises because the conditional distributions of
$(\delta_{01},\delta_{10})$ do not depend directly on $\beta$, and vice 
versa. Each term on the right hand
side of equation~\eqref{eq:POModel2} can be obtained as,
%-----------------------------
\begin{equation}
\begin{split}
	\pi(\beta^{\ast}|y^{obs}) & = \int 
	\pi(\beta^{\ast}|w,z,y^{obs}) \,
	\pi(w,z|y^{obs}) \, dw \, dz, \\
	%-----
	\pi(\delta_{j}^{\ast}|y^{obs}) 
	& = \int \pi(\delta_{j}^{\ast}| y,y^{obs})
	\, \pi(y|y^{obs}) \, dy, \qquad \textrm{where } \; j \in \{01, 10 \}.
	%-----
	%\pi(\delta_{10}^{\ast}|y^{obs}) 
	% & = \int \pi(\delta_{10}^{\ast}| y,y^{obs})
	%\, \pi(y|y^{obs}) \, dy.
\end{split}
\label{eq:CondPOModel2}
\end{equation}
%-----------------------------
Note that the density $\pi(y|y^{obs})$ is marginal with respect to
$(\beta,\delta_{01},\delta_{10})$. Consequently, the Gibbs draws of $y$
obtained from the main algorithm can be used directly, since they are not
conditional on any fixed value of
$(\beta,\delta_{01},\delta_{10})$. By disregarding the sampled values of
$(\beta,\delta_{01},\delta_{10})$, the draws of $y$ are effectively obtained
from the marginal distribution of
$y$ with respect to $(\beta,\delta_{01},\delta_{10})$. 
 
Each posterior ordinate in equation~\eqref{eq:CondPOModel2} can then be 
approximated using the Gibbs output as follows,
%-----------------------------
\begin{equation}
\begin{split}
	\hat{\pi}(\beta^{\ast}|y^{obs}) & = G^{-1} \sum_{g=1}^{G}
	\pi(\beta^{\ast}| z^{(g)}, w^{(g)},y^{obs} ), \\
	\hat{\pi}(\delta_{j}^{\ast}|y^{obs}) & = G^{-1} \sum_{g=1}^{G}
	\pi(\delta_{j}^{\ast}| y^{(g)},y^{obs} ), \qquad \textrm{where } \; j \in \{01, 10 \}.
	%\hat{\pi}(\delta_{10}^{\ast}|y^{obs}) & = G^{-1} \sum_{g=1}^{G}
	% \pi(\delta_{10}^{\ast}| y^{(g)},y^{obs} ).
	\label{eq:EstCondPOModel2}
\end{split}
\end{equation}
%-----------------------------
Substituting the density estimates from equation~\eqref{eq:EstCondPOModel2}
into equation~\eqref{eq:logBMLI} yields the following estimate of log marginal likelihood:
%-----------------------------
\begin{equation}
\begin{split}
	\ln \hat{m}(y^{obs}) & = \ln 
	f(y^{obs}|\beta^{\ast},\delta_{01}^{\ast},\delta_{10}^{\ast})
	+ \ln \Big[ \pi(\beta^{\ast}) \pi(\delta_{01}^{\ast})  
	\pi(\delta_{10}^{\ast})  \Big] \\
	& \quad - \ln \Big[ \hat{\pi}(\beta^{\ast}|y^{obs}) 
	\hat{\pi}(\delta_{01}^{\ast}|y^{obs}) 
	\hat{\pi}(\delta_{10}^{\ast}|y^{obs}) \Big],
\end{split}
\end{equation}
%-----------------------------
where the likelihood function~\eqref{eq:bqrMis-fullLike} and the prior densities for $\beta$ and ($\delta_{01},\delta_{10}$) are all evaluated at a high-density point.

%-------------------------------------------------------------------------------
\section{Simulation}\label{sec:Simulations}
%-------------------------------------------------------------------------------

We conduct a series of simulation studies to evaluate the performance of the proposed model under different combinations of prior effective sample size ($n_{\text{pess}}$), misclassification rates, and prior variance. First, we set $n_{\text{pess}}$ to 30, 50, and 100 and use these values to construct the prior distributions for the false negative ($\delta_{01}$) and false positive ($\delta_{10}$) rates. Second, we consider three levels of misclassification by specifying the following combinations of false negative and false positive rates: $(\delta_{01}=0.4, \delta_{10}=0.2)$, $(\delta_{01}=0.3, \delta_{10}=0.2)$, and $(\delta_{01}=0.2, \delta_{10}=0.1)$. Third, we examine the sensitivity of the results to prior dispersion by varying the prior variance of the regression coefficients, with $B_{0}=5I$ and $B_{0}=10I$, where $I$ denotes the identity matrix.

For each scenario, we generate 100 synthetic data sets, each with a sample size of $n=1000$. The covariate vector is three-dimensional, where the first element is set to 1 and the remaining elements are simulated from the standard normal distribution. The latent continuous variable is generated as: $
z_i = x'_{i}\beta + \varepsilon_i$, where $\beta = (0,1,-0.5)'$ and $\varepsilon_i \sim \text{AL}(0,1,p)$ with location 0, scale 1, and quantile parameter $p$. The true binary response is defined as $ y_i = I(z_i > 0)$. To introduce misclassification, the observed binary response $y^{obs}_i$ is generated as
\[
y^{obs}_i  \sim
\begin{cases}
\text{Bernoulli}(1-\delta_{01}), & \text{if } y_{i} = 1, \\
\text{Bernoulli}(\delta_{10}), & \text{if } y_{i} = 0.
\end{cases}
\]

We envision that the priors for the misclassification parameters will arise either via validation data or expert elicited information that is converted into Beta distributions. To account for the variability that would accompany this process, we simulate prior effective sample size, $n_{pess}=30, 50, 100$, for both the false negative and false positive probabilities i.e., $t_{10} \sim Bin(n_{\text{pess}} , \delta_{10})$, and $t_{01} \sim Bin(n_{\text{pess}}, \delta_{01})$, where $Bin(\cdot,\cdot)$ denotes a Binomial distribution. These quantities are converted directly to Beta prior distributions: $
\delta_{10} \sim B(t_{10} + 1, n_{\text{pess}} - t_{10} + 1)$ and 
$\delta_{01} \sim B(t_{01} + 1, n_{\text{pess}} - t_{01} + 1)$. Thus, the priors are ``on average" correct, but will typically not be centered exactly on the truth for any individual simulated data set.

We implement two independent Gibbs chains with 10,000 iterations per chain after a burn-in of 5,000 iterations. Convergence is assessed by visual inspection of the trace plots and the Gelman–Rubin potential scale reduction factor. The trace plots show stable behavior with good mixing across chains and no evidence of nonstationarity or lack of convergence. A representative trace plot for the 50-th quantile, with $n_{pess}=30$, $(\delta_{01}=0.4, \delta_{10}=0.2)$ and $B_{0}=10*I$ is presented in Figure~\ref{mcmc} (\ref{app:trace}). The Gelman–Rubin statistics for all monitored parameters are close to 1, indicating negligible between-chain variability relative to within-chain variability and providing evidence of convergence.

Tables~\ref{Table:Bias-Sim}, \ref{Table:MSE-Sim}, and \ref{Table:CR-Sim} present the average biases, mean squared errors (MSE), and coverage rates, respectively, of the posterior estimates of regression coefficients from the binary quantile model without misclassification (BQR) and the proposed binary quantile model with misclassification (BQRM) model under ESS of 30 and a prior variances of $\beta$ equal to 10. The remaining cases are summarized in Tables~\ref{Table:Case1_AB}-\ref{Table:Case3_Coverage} (\ref{app:sim}). Overall, the results demonstrate that explicitly accounting for misclassification substantially improves estimation accuracy and uncertainty quantification.

%------------------------------------------------------------------------
\begin{table}[!t]
\centering
\caption{Average biases of posterior estimates of regression coefficients from the two models when prior effective sample size, $n_{\text{pess}}$ = 30 and $B_{0}=10I_3$.} 
\label{Table:Bias-Sim}
\begin{tabular}{ccc rrr rrr}
\toprule
 &  &  & \multicolumn{3}{c}{BQR} & \multicolumn{3}{c}{BQRM} \\
\cmidrule(lr){4-6} \cmidrule(lr){7-9}
%  \midrule
  $\delta_{01}$ & $\delta_{10}$ & $p$ & $\beta_{1}$ & $\beta_{2}$ & $\beta_{3}$ & $\beta_{1}$ & $\beta_{2}$ & $\beta_{3}$ \\ 
  \midrule

0.4 & 0.2 & $0.10$ & $-15.09$ & $-0.56$ & $0.25$ & $-12.27$ & $-0.49$ & $0.24$ \\ 
   &  & 0.25 & $-2.49$ & $-0.63$ & $0.29$ & $-1.93$ & $0.02$  & $-0.06$ \\ 
   &  & 0.50 & $-0.45$ & $-0.69$ & $0.36$ & $-0.03$ & $-0.39$ & $0.21$ \\ 
   &  & 0.75 & $0.66$  & $-0.66$ & $0.32$ & $1.64$  & $0.06$  & $-0.06$ \\ 
   &  & 0.90 & $8.73$  & $-0.49$ & $0.23$ & $13.22$ & $-0.43$ & $0.24$ \\ 
        \midrule
0.3 & 0.2 & $0.10$ & $-12.38$ & $-0.54$ & $0.27$ & $-12.77$ & $-0.39$ & $0.19$ \\ 
   &  & 0.25 & $-1.45$ & $-0.62$ & $0.31$ & $-1.46$ & $0.22$ & $-0.12$ \\ 
   &  & 0.50 & $-0.23$ & $-0.63$ & $0.32$ & $-0.02$ & $-0.26$ & $0.13$ \\ 
   &  & 0.75 & $0.92$ & $-0.56$ & $0.27$ & $1.98$ & $0.38$ & $-0.20$ \\ 
   &  & 0.90 & $8.62$ & $-0.39$ & $0.25$ & $12.96$ & $-0.21$ & $0.13$ \\
    \midrule
0.2 & 0.1 & $0.10$ & $-9.46$ & $-0.21$ & $0.10$ & $-13.37$ & $0.14$ & $-0.09$ \\ 
   &  & 0.25 & $-1.45$ & $-0.36$ & $0.18$ & $-2.36$ & $0.64$ & $-0.33$ \\ 
   &  & 0.50 & $-0.24$ & $-0.44$ & $0.21$ & $0.00$ & $-0.10$ & $0.04$ \\ 
   &  & 0.75 & $0.48$ & $-0.37$ & $0.18$ & $1.80$ & $0.67$ & $-0.34$ \\ 
   &  & 0.90 & $6.19$ & $0.01$ & $-0.01$ & $15.07$ & $0.43$ & $-0.29$ \\
   \bottomrule
\end{tabular}
\end{table}
%------------------------------------------------------------------------

%\sout{Table~\ref{Table:Bias-Sim} shows that the proposed model yields estimates that are much closer to the true parameter values across quantiles and misclassification settings. The naive model, by contrast, exhibits noticeable bias, particularly under higher levels of misclassification. This bias attenuation pattern reflects the systematic distortion caused by ignoring outcome misclassification. Overall, the proposed model demonstrates superior estimation accuracy and robustness across varying degrees of misclassification.}

Table~\ref{Table:Bias-Sim} shows that accounting for misclassification substantially reduces bias for several regression coefficients, particularly at the central quantiles. The improvement is most evident for $\beta_2$ and $\beta_3$, for which BQRM generally produces estimates closer to their true values than the BQR model. For example, when $(\delta_{01},\delta_{10})=(0.4,0.2)$, the biases at $p=0.50$ decline from $(-0.69,0.36)$ under BQR to $(-0.39,0.21)$ under BQRM for $(\beta_2,\beta_3)$. In contrast, the bias of $\beta_1$ remains substantial at the extreme quantiles, with BQRM yielding larger absolute biases at $p=0.90$ across all three misclassification settings. These results indicate that accounting for misclassification improves estimation accuracy in the central portion of the conditional distribution, as reflected in both the lower bias and MSE of BQRM, while tail estimation, particularly for $\beta_1$, remains more sensitive and challenging.

%------------------------------------------------------------------------
\begin{table}[!t]
\centering
\caption{Mean squared errors (MSE) of posterior estimates of regression coefficients from the two models when prior effective sample size, $n_{\text{pess}}$ = 30 and $B_{0}=10I_3$.} 
\label{Table:MSE-Sim}
\begin{tabular}{ccc ccc ccc}
\toprule
 &  &  & \multicolumn{3}{c}{BQR} & \multicolumn{3}{c}{BQRM} \\
\cmidrule(lr){4-6} \cmidrule(lr){7-9}
%  \midrule
  $\delta_{01}$ & $\delta_{10}$ & $p$ & $\beta_{1}$ & $\beta_{2}$ & $\beta_{3}$ & $\beta_{1}$ & $\beta_{2}$ & $\beta_{3}$ \\ 
  \midrule
0.4 & 0.2 & 0.10 & 421.98 & 0.55 & 0.19 & 186.67 & 0.48 & 0.26 \\ 
   &  & 0.25 & 9.84 & 0.45 & 0.11 & 8.09 & 0.17 & 0.09 \\ 
   &  & 0.50 & 0.21 & 0.49 & 0.13 & 0.02 & 0.19 & 0.07 \\ 
   &  & 0.75 & 0.72 & 0.46 & 0.11 & 4.83 & 0.18 & 0.11 \\ 
   &  & 0.90 & 95.92 & 0.35 & 0.14 & 197.07 & 0.43 & 0.21 \\ 
    \midrule
0.3 & 0.2 & 0.10 & 243.68 & 0.45 & 0.16 & 206.88 & 0.42 & 0.23 \\ 
   &  & 0.25 & 3.74 & 0.43 & 0.11 & 6.03 & 0.25 & 0.12 \\ 
   &  & 0.50 & 0.06 & 0.41 & 0.11 & 0.04 & 0.13 & 0.05 \\ 
   &  & 0.75 & 1.22 & 0.35 & 0.09 & 6.59 & 0.38 & 0.16 \\ 
   &  & 0.90 & 99.69 & 0.30 & 0.16 & 202.90 & 0.20 & 0.26 \\
 
   \midrule
0.2 & 0.1 & 0.10 & 129.14 & 0.29 & 0.13 & 241.20 & 0.28 & 0.21 \\ 
   &  & 0.25 & 3.06 & 0.20 & 0.06 & 9.88 & 0.68 & 0.25 \\ 
   &  & 0.50 & 0.06 & 0.21 & 0.05 & 0.07 & 0.10 & 0.03 \\ 
   &  & 0.75 & 0.42 & 0.18 & 0.05 & 6.22 & 0.77 & 0.25 \\ 
   &  & 0.90 & 47.16 & 0.16 & 0.11 & 256.05 & 0.43 & 0.35 \\
   \bottomrule
\end{tabular}
\end{table}
%------------------------------------------------------------------------

%\sout{In Table~\ref{Table:MSE-Sim}, the mean squared errors (MSE) demonstrate that the proposed BQRM model consistently achieves lower MSEs than the BQR model across all scenarios. In particular, the proposed model substantially reduces MSEs at the central quantiles when misclassification is severe.}

Table~\ref{Table:MSE-Sim} shows that accounting for misclassification generally improves estimation, particularly at the central quantiles and for $\beta_2$ and $\beta_3$. The gains are especially pronounced at the median; for example, when $(\delta_{01},\delta_{10})=(0.4,0.2)$, the MSEs at $p=0.50$ decline from $(0.21,0.49,0.13)$ under BQR to $(0.02,0.19,0.07)$ under BQRM. However, the improvement is not uniform across coefficients and quantiles. In particular, 
MSEs for $\beta_{1}$ are substantially larger at $p=0.90$ in all three misclassification scenarios. To sum it up, the results suggest that explicitly modeling misclassification provides the greatest benefits for estimation in the central portion of the conditional distribution, while tail estimation remains more challenging.

%------------------------------------------------------------------------
\begin{table}[!t]
\centering
\caption{Coverage rates of posterior estimates of regression coefficients from the two models when prior effective sample size, $n_{\text{pess}}$ = 30 and $B_{0}=10I_3$.} 
\label{Table:CR-Sim}
\begin{tabular}{ccc ccc ccc}
\toprule
 &  &  & \multicolumn{3}{c}{BQR} & \multicolumn{3}{c}{BQRM} \\
\cmidrule(lr){4-6} \cmidrule(lr){7-9}
%  \midrule
  $\delta_{01}$ & $\delta_{10}$ & $p$ & $\beta_{1}$ & $\beta_{2}$ & $\beta_{3}$ & $\beta_{1}$ & $\beta_{2}$ & $\beta_{3}$ \\ 
  \midrule
    
0.4 & 0.2 & 0.10 & 0.00 & 0.44 & 0.51 & 1.00 & 1.00 & 1.00 \\ 
   &  & 0.25 & 0.00 & 0.29 & 0.45 & 1.00 & 1.00 & 1.00 \\ 
   &  & 0.50 & 0.00 & 0.00 & 0.15 & 1.00 & 0.99 & 0.96 \\ 
   &  & 0.75 & 0.96 & 0.33 & 0.50 & 1.00 & 1.00 & 1.00 \\ 
   &  & 0.90 & 0.19 & 0.73 & 0.74 & 1.00 & 1.00 & 1.00 \\ 
     \midrule
0.3 & 0.2 & 0.10 & 0.25 & 0.56 & 0.57 & 1.00 & 1.00 & 1.00 \\ 
   &  & 0.25 & 0.02 & 0.37 & 0.40 & 1.00 & 1.00 & 1.00 \\ 
   &  & 0.50 & 0.20 & 0.06 & 0.37 & 1.00 & 0.99 & 0.97 \\ 
   &  & 0.75 & 0.91 & 0.56 & 0.62 & 1.00 & 1.00 & 1.00 \\ 
   &  & 0.90 & 0.24 & 0.73 & 0.71 & 1.00 & 1.00 & 1.00 \\ 
    \midrule
   0.2 & 0.1 & 0.10 & 0.29 & 0.72 & 0.68 & 0.99 & 1.00 & 1.00 \\ 
   &  & 0.25 & 0.34 & 0.67 & 0.65 & 1.00 & 1.00 & 1.00 \\ 
   &  & 0.50 & 0.17 & 0.60 & 0.82 & 1.00 & 1.00 & 1.00 \\ 
   &  & 0.75 & 1.00 & 0.68 & 0.76 & 1.00 & 1.00 & 0.99 \\ 
   &  & 0.90 & 0.77 & 0.89 & 0.85 & 1.00 & 1.00 & 1.00 \\ 
   \bottomrule
\end{tabular}
\end{table}
%------------------------------------------------------------------------

%Table~\ref{Table:CR-Sim} shows that the proposed model achieves coverage rates at or above the nominal 95\% level across most parameter settings, whereas the naive model tends to underestimate uncertainty, resulting in coverage rates substantially below nominal levels, particularly at the central quantile. The proposed model also provides consistent coverage across different levels of misclassification, while the naive model presents improvement in coverage as the degree of misclassification decreases. These findings indicate that accounting misclassification enhances uncertainty quantification and prevents false positive conclusion in hypothesis testing. The above nominal coverage for the parameters in the proposed model is not surprising given the additional uncertainty from accounting for the misclassification. \citet{GustafsonBook-2003} provides a thorough discussion of this phenomenon.

Table~\ref{Table:CR-Sim} shows that the proposed BQRM model achieves coverage rates at or above 95\% level across all parameter settings and quantiles, whereas the BQR model often exhibits substantial undercoverage, particularly for $\beta_2$ and $\beta_3$ and at the central quantiles. The coverage of BQRM remains stable across different levels of misclassification, whereas BQR coverage generally improves as misclassification decreases. These results indicate that accounting for misclassification substantially improves uncertainty quantification and yields more reliable posterior inference. The slightly conservative coverage of BQRM is expected because the model explicitly incorporates the additional uncertainty arising from unknown misclassification rates; see \citet{GustafsonBook-2003} for a detailed discussion of this phenomenon.

%------------------------------------------------------------------------------
\begin{figure}[!b]
    \centering
    \includegraphics[width=0.8\textwidth]{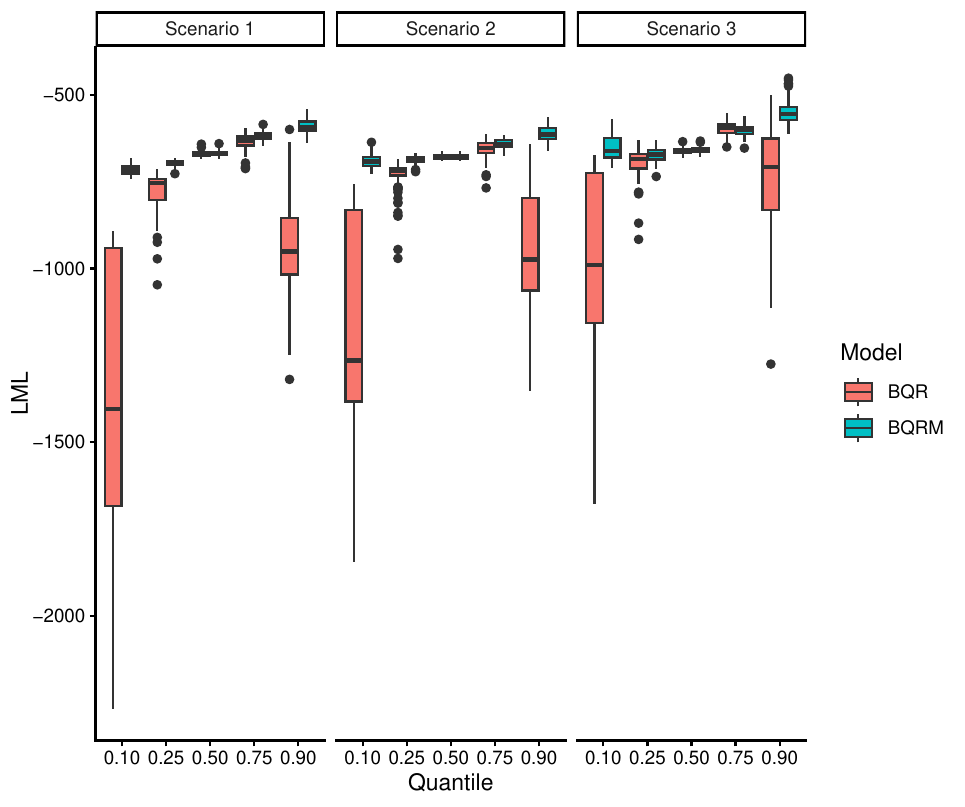}
    \caption{The log marginal likelihoods (LMLs) for the two models when prior effective sample size $n_{pess}=30$ and $B_0 =10 I_3$. Scenario~1: $(\delta_{01}=0.4,\delta_{10}=0.2)$, Scenario~2: $(\delta_{01}=0.3,\delta_{10}=0.2)$, and Scenario~3: $(\delta_{01}=0.2,\delta_{10}=0.1)$.}
    \label{fig:logML}
\end{figure}
%------------------------------------------------------------------------------

Finally, we compute the marginal likelihood for each simulation and use it to compare the BQR and BQRM models at each quantile. Across the simulation settings, the BQRM model generally yields higher log marginal likelihoods than the BQR model, indicating better model fit. Figure~\ref{fig:logML} illustrates this comparison using box plots of the log marginal likelihoods across quantiles and misclassification rates for $n_{\text{pess}}=30$ and $B_{0}=10I$. The BQRM model exhibits higher average log marginal likelihoods and, in most cases, a narrower interquartile range than BQR, indicating more favorable and stable model fit across simulations. The difference is particularly pronounced at the lower and upper quantiles, whereas the two models exhibit more similar fit at the median.

%------------------------------------------------------------------------------
\section{Application} \label{sec:Application}
%------------------------------------------------------------------------------

In this section, we apply the BQR and the proposed BQRM models developed in Section~\ref{sec:Method} to analyze women's self-reports of spousal violence based on the NFHS data described in Section~\ref{sec:Data}. Our approach is Bayesian, so we need to specify prior distributions for the regression coefficients $\beta$ and the misclassification parameters $(\delta_{01},\delta_{10})$. While relatively diffuse priors may be appropriate for $\beta$, the specification of priors for $(\delta_{01},\delta_{10})$ should ideally incorporate available expert knowledge about the prevalence and direction of misreporting.

For the regression coefficients, we adopt Gaussian priors, $\beta \sim N(0, 10 I_k)$, where \(I_k\) denotes the $(k \times k)$ identity matrix and $k$ is the number of regression parameters.  This prior specification is used for both BQR and BQRM models. For the misclassification probabilities $(\delta_{01}, \delta_{10})$, we specify priors based on external evidence from \cite{RABIN2009439}, who study domestic violence reporting using a multiple-measurement approach. Specifically, we let,
%------------------
\begin{equation*}
\delta_{01} \sim \text{Beta}(7.6, 5), \quad \delta_{10} \sim \text{Beta}(9.7, 165.7).
\end{equation*}
%-------------------
These priors imply $E(\delta_{01}) = 0.6032$ and $E(\delta_{10}) = 0.0553$, with corresponding variances of $0.0176$ and $0.0003$, respectively. Consequently, the prior for the false negative rate $\delta_{01}$ is centered at a substantially larger value than that for the false positive rate $\delta_{10}$, which helps ensure parameter identification through an ordering restriction \citep{Wang-etal-2026}. Moreover, the prior on \(\delta_{01}\) is considerably more dispersed, allowing greater uncertainty about the extent of underreporting while maintaining a relatively tight prior on false positive reporting. This specification is consistent with a large body of empirical evidence indicating that victims are more likely to conceal or underreport experiences of domestic violence than to falsely report such events. 

We generate two independent Gibbs chains, each consisting of 100,000 Gibbs iterations. In each chain, the first 50,000 iterations are discarded as burn-in, and results are reported based on the post burn-in Gibbs sample. Trace plots (not shown) and Gelman-Rubin's diagnostic suggests convergence of Gibbs chains, and are similar to those of the simulation studies. 

The posterior means and 95\% credible intervals for the parameters in the BQR and BQRM models are reported in Table~\ref{Table:AppPostSum}. Our first observation from the table is that the posterior mean of the false negative rate, $\delta_{01}$, consistently exceeds that of the false positive rate, $\delta_{10}$, across all quantiles. This finding is consistent with empirical evidence that underreporting of spousal violence is substantially more prevalent than false accusations \citep{Chin-etal-2017, Polettini-etal-2024, Arima-etal-2026}.

%---------------------------------------------------------------------
\begin{table}[t!]
\centering
\caption{Posterior means and 95\% credible intervals (CrI).}
\label{Table:AppPostSum}
\setlength{\tabcolsep}{6pt}

\begin{tabular}{l l S S S S S S}
\toprule
& & \multicolumn{3}{c}{BQR}
    & \multicolumn{3}{c}{BQRM} \\
\cmidrule(lr){3-5}\cmidrule(lr){6-8}

Quantile & Variable
& \multicolumn{1}{c}{Mean}
& \multicolumn{2}{c}{95\% CrI}
& \multicolumn{1}{c}{Mean}
& \multicolumn{2}{c}{95\% CrI} \\
\cmidrule(lr){4-5}\cmidrule(lr){7-8}

& & & \multicolumn{1}{c}{Lower} & \multicolumn{1}{c}{Upper}
& & \multicolumn{1}{c}{Lower} & \multicolumn{1}{c}{Upper} \\
\midrule

0.25 & Intercept  & -3.00  & -4.72 & -1.82 & -0.44  & -6.78 & 1.26 \\
     & fage       & -0.25  & -0.38 & -0.15 & -0.34  & -1.18 & -0.02 \\
     & fwork      &  0.19  &  0.05 &  0.37 &  0.32  & -0.11 & 1.19 \\
     & meduc      & -0.32  & -0.49 & -0.20 & -0.54  & -1.74 & -0.05 \\
     & wealth     & -0.70  & -1.01 & -0.48 & -1.33  & -3.89 & -0.31 \\
     & nchildren  &  0.24  &  0.14 &  0.36 &  0.54  &  0.10 & 1.54 \\
     & remarriage &  0.19  & -0.20 &  0.63 &  1.14  & -0.59 & 4.06 \\
     & polyg      &  0.90  &  0.41 &  1.50 &  3.21  &  0.54 & 7.47 \\
     & nwomen     & -0.05  & -0.12 &  0.02 & -0.05  & -0.32 & 0.14 \\
     & $\delta_{01}$ & {\text{--}} & {\text{--}} & {\text{--}}
                       & 0.75 & 0.55 & 0.81 \\
     & $\delta_{10}$ & {\text{--}} & {\text{--}} & {\text{--}}
                       & 0.01 & 0.00 & 0.06 \\

\midrule

0.5  & Intercept  & -1.95  & -2.99 & -1.22 & -0.42  & -1.02 & -0.07 \\
     & fage       & -0.21  & -0.32 & -0.12 & -0.23  & -0.52 & -0.05 \\
     & fwork      &  0.17  &  0.05 &  0.33 &  0.22  &  0.02 & 0.59 \\
     & meduc      & -0.27  & -0.40 & -0.16 & -0.30  & -0.72 & -0.06 \\
     & wealth     & -0.58  & -0.83 & -0.39 & -0.74  & -1.73 & -0.19 \\
     & nchildren  &  0.20  &  0.12 &  0.30 &  0.27  &  0.06 & 0.65 \\
     & remarriage &  0.17  & -0.17 &  0.56 &  0.37  & -0.27 & 1.33 \\
     & polyg      &  0.77  &  0.32 &  1.39 &  1.45  &  0.29 & 3.42 \\
     & nwomen     & -0.04  & -0.10 &  0.02 & -0.04  & -0.15 & 0.04 \\
     & $\delta_{01}$ & {\text{--}} & {\text{--}} & {\text{--}}
                       & 0.62 & 0.58 & 0.66 \\
     & $\delta_{10}$ & {\text{--}} & {\text{--}} & {\text{--}}
                       & 0.01 & 0.00 & 0.02 \\

\midrule

0.75 & Intercept  & -0.66  & -1.00 &  0.69 &  1.62  & -1.07 & 12.60 \\
     & fage       & -0.24  & -0.50 & -0.11 & -0.49  & -1.46 & -0.12 \\
     & fwork      &  0.21  &  0.05 &  0.54 &  0.44  &  0.03 & 1.48 \\
     & meduc      & -0.28  & -0.59 & -0.13 & -0.62  & -2.03 & -0.14 \\
     & wealth     & -0.65  & -1.42 & -0.30 & -1.46  & -4.30 & -0.33 \\
     & nchildren  &  0.23  &  0.10 &  0.50 &  0.54  &  0.11 & 1.60 \\
     & remarriage &  0.23  & -0.18 &  0.86 &  0.58  & -0.74 & 2.89 \\
     & polyg      &  1.00  &  0.27 &  2.45 &  2.38  &  0.32 & 7.16 \\
     & nwomen     & -0.04  & -0.14 &  0.01 & -0.09  & -0.41 & 0.07 \\
     & $\delta_{01}$ & {\text{--}} & {\text{--}} & {\text{--}}
                       & 0.46 & 0.03 & 0.82 \\
     & $\delta_{10}$ & {\text{--}} & {\text{--}} & {\text{--}}
                       & 0.02 & 0.00 & 0.05 \\
\bottomrule
\end{tabular}
\end{table}
%------------------------------------------------------------------------------
 
The posterior mean of $\delta_{01}$ is 0.62 at the median and ranges from 0.46 to 0.75 across the 10th to 90th quantiles (see Table~\ref{Table:AppPostResSupp} for the extreme quantiles). The relatively narrow 95\% credible intervals lend credibility to these estimates. Although the estimated false negative rates may appear high, they are broadly consistent with previous findings in the literature \citep{Chin-etal-2017, Polettini-etal-2024,Arima-etal-2026}. In contrast, the posterior mean of the false positive rate, $\delta_{10}$, is only 0.01 at the median and ranges from 0.01 to 0.03 across quantiles, with similarly narrow credible intervals. Overall, these results indicate substantial underreporting and underscore the importance of accounting for response misclassification when modeling women's reports of spousal violence.

Second, we observe from Table~\ref{Table:AppPostSum} that regression coefficients credibly different from zero (i.e., whose 95\% credible intervals exclude zero) in the BQR model generally remain so in the BQRM model. The only exception is women’s employment status (\texttt{fwork}), which is credibly different from zero at the 25th quantile under BQR but not under BQRM. At the extreme quantiles (10th and 90th; see Table~\ref{Table:AppPostResSupp} in \ref{app:india}), however, most variables that are credibly different from zero under BQR are no longer distinguishable from zero under BQRM. The exception is the wealth index (\texttt{wealth}), which exhibits a consistently negative association with the propensity to report spousal violence across quantiles. Another notable feature, evident from Table~\ref{Table:AppPostSum}, is that credible intervals under the BQRM model are generally wider than those under the BQR model. This pattern is typical of misclassification models and reflects the additional uncertainty arising from the estimation of misclassification parameters.

On the direction of covariate effect, we observe from Table~\ref{Table:AppPostSum} that in the BQRM model and across three quartiles, the propensity to report spousal violence is positively associated with women's employment (\texttt{fwork}), number of children (\texttt{nchildren}), and being in a polygynous relationship (\texttt{polyg}). Among these variables, women's employment (\texttt{fwork}) has received considerable attention in the literature. Studies such as \citet{Dutta-etal-2016} and \citet{Mishra-etal-2024} find that working women face a higher risk of spousal violence. This relationship has been attributed either to male backlash \citep{Krishnan-etal-2010,Chin-2012}, whereby men respond to women's employment with increased violence, or to reverse causality, whereby women experiencing violence seek employment to attain financial independence \citep{Bhattacharya-2015}. While the coefficient for women's employment is consistent with this literature, the BQRM results, however, reveal greater heterogeneity: employment is not credibly associated with reporting at the lower quantile but becomes relevant at the median and third quartile. Thus, accounting for misclassification suggests that the role of employment depends on the underlying reporting propensity. The positive associations with the number of children (\texttt{nchildren}) and polygynous relationships (\texttt{polyg}) may, respectively, reflect greater household resource constraints and reduced bargaining power, consistent with the evidence in \citet{Mishra-etal-2024}.

We next consider the variables negatively associated with reporting spousal violence. The wealth index (\texttt{wealth}) is particularly noteworthy because its negative association persists not only between the 25th and 75th quantiles but also at the extreme quantiles (see Table~\ref{Table:AppPostResSupp}). Although this relationship has been documented using conditional-mean models, such as logistic regression \citep{Mishra-etal-2024}, it has not, to our knowledge, been examined across the distribution of reporting propensity, either with or without accounting for misclassification. Our results therefore extend the existing evidence by revealing heterogeneity in this relationship across the distribution. The negative association may reflect greater reluctance among women from wealthier households to report violence, possibly because of concerns about social status and family reputation, as well as greater household resources to manage or withstand the consequences of legal intervention. These factors may make women in more affluent households more tolerant of, or less willing to disclose, spousal violence.

Women's age (\texttt{fage}) and husband's education (\texttt{meduc}) also exhibit negative associations with reporting, consistent with previous empirical evidence \citep{Dutta-etal-2016, Mishra-etal-2024} and theoretical expectations. Older women may be less inclined to report violence because of greater concerns about family reputation and social standing, while longer marital duration may contribute to the normalization or tolerance of abusive behavior. Higher levels of husband's education, in contrast, may be associated with greater social awareness and more egalitarian attitudes, potentially reducing both the incidence of spousal violence and the need to report it \citep{Bhattacharya-etal-2011}.

To quantify the magnitude of covariate effects and their variation across the distribution, we compute the average covariate effects described in Section~\ref{sec:CovEff} and report them in Table~\ref{Table:CovEffect}, restricting attention to statistically credible variables. For a 10-year increase in women's age, the estimated effects under BQRM range from $-0.036$ at the 25th quantile to $-0.048$ at the 75th quantile, compared with approximately $-0.03$ under BQR. A 4-year increase in husband's education has relatively stable effects across quantiles, but the magnitude is larger under BQRM (about $-0.035$) than under BQR ($-0.022$). For a one-standard-deviation increase in wealth, the estimates under BQRM are $-0.110$, $-0.102$, and $-0.106$ at the 25th, 50th, and 75th quantiles, respectively, compared with $-0.060$, $-0.054$, and $-0.061$ under BQR. For women's employment, the estimates increase from $0.032$ at the median to $0.036$ at the 75th quantile, compared with $0.018$ and $0.023$, respectively, under BQR. The effect of an additional child is $0.032$ at the 25th quantile and $0.029$ at both the 50th and 75th quantiles, compared with $0.016$, $0.015$, and $0.017$ under BQR. The largest difference is observed for polygamy: the estimates under BQRM are $0.189$, $0.217$, and $0.186$ at the three quartiles compared with $0.092$, $0.095$, and $0.118$ under BQR. At the extreme quantiles, fewer covariates are credible under BQRM, with wealth remaining credible at the 10th and 90th quantiles, where the corresponding estimates are $-0.073$ and $-0.049$, respectively. Overall, the covariate effects tend to be larger in magnitude under BQRM than under BQR.

%------------------------------------------------------------------------------
\begin{table}[!t]
\centering
\caption{Average covariate effects for statistically credible variables across quantiles.}
\label{Table:CovEffect}
\small
\setlength{\tabcolsep}{6pt}
\begin{tabular}{ccrrrrrrr}
\toprule
$p$ & Model & \texttt{fage} & \texttt{fwork} & \texttt{medu} & \texttt{wealth} & \texttt{nchild} & \texttt{polyg} \\
\midrule
0.10 & BQR  & $-0.025$ & $0.016$ & $-0.020$ & $-0.053$  & $0.013$ & $0.075$ \\
     & BQRM &        &       &        & $-0.073$ &        &       \\
\midrule
0.25 & BQR  & $-0.029$ & $0.018$ & $-0.023$ & $-0.060$ & $0.016$ & $0.092$ \\
     & BQRM & $-0.036$ &       & $-0.034$ & $-0.110$ & $0.032$ & $0.189$ \\
\midrule
0.50 & BQR  & $-0.027$ & $0.018$ & $-0.021$ & $-0.054$ & $0.015$ & $0.095$ \\
     & BQRM & $-0.043$ & $0.032$ & $-0.034$ & $-0.102$ & $0.029$ & $0.217$ \\
\midrule
0.75 & BQR  & $-0.030$ & $0.023$ & $-0.022$ & $-0.061$ & $0.017$ & $0.118$ \\
     & BQRM & $-0.048$ & $0.036$ & $-0.036$ & $-0.106$ & $0.029$ & $0.186$ \\
\midrule
0.90 & BQR  & $-0.033$ & $0.025$ & $-0.023$ & $-0.075$ & $0.018$ & $0.107$ \\
     & BQRM &        &       &        & $-0.049$ &        &       \\
\bottomrule
\end{tabular}
\end{table}
%------------------------------------------------------------------------------

Finally, we compare the model support for the BQR and BQRM models of women's self-reported spousal violence using the marginal likelihoods computed as described in Section~\ref{sec:ML}. At the considered quantiles, the log marginal likelihoods for the BQR models are $-14917$, $-10632$, $-8563$, $-8375$, and $-12114$, while the corresponding values for the BQRM models are $-8464$, $-8548$, $-8434$, $-8933$, and $-10467$. Since a larger log marginal likelihood indicates stronger support for a model given the observed data, these results provide substantially stronger support for the BQRM model at the 10th, 25th, 50th, and 90th quantiles. At the 75th quantile, however, the BQR model receives greater support. The magnitude of the differences in log marginal likelihood also varies considerably across quantiles, with particularly strong support for BQRM at the lower quantiles and at the 90th quantile, but weaker support around the median and greater support for BQR at the 75th quantile. Overall, the results suggest that accounting for misclassification provides substantially greater model support at most quantiles, implying that misclassification is consequential and that treating self-reported violence as the true outcome does not adequately capture the observed reporting patterns.

%\begin{figure}[htbp]
%    \centering
%    \includegraphics[width=0.8\textwidth]{ML.pdf}
%    \caption{The log marginal likelihoods (LMLs) for the two models across five quantile levels. }
%    \label{fig:myfigure}
%\end{figure}
%------------------------------------------------------------------------------
\section{Conclusions and discussion}\label{sec:Conclusion}
%------------------------------------------------------------------------------

We introduce a Bayesian quantile regression framework for binary responses subject to misclassification, grounded in a latent variable representation in which misclassification probabilities are treated as unknown parameters rather than pre-specified constants. This formulation fundamentally relaxes the restrictive assumption of fixed false negative and false positive rates--equivalently, fixed sensitivity and specificity--and permits coherent propagation of misclassification uncertainty through all stages of posterior inference, yielding more honest and calibrated inference on the regression coefficients. The latent variable structure further affords tractable model construction and efficient posterior computation via a novel partially collapsed Gibbs sampler, which strategically exploits conditional independencies to accelerate mixing and reduce autocorrelation in the Markov chain. We also attend to post-estimation constructs and discuss the computation of covariate effects from the Gibbs draws at user-specified quantiles, and derive the marginal likelihood using the method of \citet{Chib-1995}, which provides a principled basis for Bayesian model comparison across specifications differing in covariates and misclassification structure, or both.

Simulation studies indicate that failure to account for outcome misclassification can substantially bias quantile-specific effect estimates and distort inference, particularly under moderate to high misclassification. In contrast, the proposed framework achieves superior estimation accuracy and more reliable uncertainty quantification, and better model support as given by the log marginal likelihood. The real-data application on women’s self-reporting of spousal violence yields three main findings. First, estimated false negative rates systematically exceed false positive rates across quantiles, confirming that underreporting of spousal violence is far more prevalent than overreporting, a well-documented phenomenon in spousal violence literature. Second, accounting for misclassification alters the determinants of self-reported spousal violence. For example, female employment status appears positively associated with spousal violence reporting in models that ignore misclassification. However, once misclassification is properly accounted, this association becomes weaker and is negligible at lower quantiles.  In contrast, household wealth retains a consistently strong and negative association with spousal violence reporting across all quantiles, proving robust to misclassification adjustment and pointing to a stable socioeconomic gradient in women's vulnerability and disclosure behavior. Third, model comparison based on marginal likelihood favors the binary quantile model with misclassification at all quantiles except the 75th (third quartile). This provides overall support for accounting for misclassification and highlights its importance when modeling self-reported spousal violence.

The binary quantile regression framework with misclassification proposed here can be developed further in at least two different directions. First, throughout this work we assumed non-differential misclassification, implying that false negative and false positive rates remain constant across all covariate profiles. In real-world settings, however, misclassification may be differential, as classification accuracy may vary systematically with individual characteristics such as education, age, or cultural context. For instance, in the spousal violence application, more educated respondents may exhibit higher reporting accuracy, inducing covariate-dependent misclassification probabilities. Similarly, other socially less desirable behavior such as reporting marijuana usage \citep{Wang-etal-2026} may be systematically related to covariates such as age and race. Consequently, quantile regression with differential misclassification presents an important avenue for future research.

Second, we often encounter high-dimensional settings in real-world applications, where a large number of potentially correlated covariates are collected. In these cases, high dimensionality not only increases computational burden but also exacerbates issues of overfitting, multicollinearity, and instability in posterior inference, especially when outcome misclassification is present. Ignoring misclassification in high-dimensional models may further inflate false discoveries or obscure truly important predictors. Therefore, scalable Bayesian shrinkage methods, such as spike-and-slab priors, are particularly promising, as they facilitate simultaneous variable selection, regularization and proper uncertainty propagation from the misclassification component. Developing computationally efficient algorithms tailored to high-dimensional misclassification model, along with establishing rigorous theoretical properties, remain an important direction for future research.

\section*{Conflicts of interest}
The authors declare that they have no competing interests.

\section*{Funding}
This research received no external funding.

\section*{Data availability}
The data used in this study are drawn from the Demographic and Health Surveys (DHS) Program. These data are publicly available upon registration at \url{https://dhsprogram.com/}.
%------------------------------ Bibliography ---------------------------------
\clearpage \pagebreak
%\nocite{*}
\pdfbookmark[1]{References}{unnumbered} % This makes References appear as a bookmark in pdf
%\section*{References}

%\bibliographystyle{abbrvnat}
\bibliographystyle{jasa}
\bibliography{BQRMisBinaryData}

\clearpage
%------------------------------------------------------------------------------
% -------------------- Beginning of Appendices -------------------------------
%------------------------------------------------------------------------------

\newpage
\appendix
\renewcommand\thesection{Appendix \Alph{section}}
%\setcounter{secnumdepth}{-1}    % This code removes the numbering of the 
%%appendix

%------------------------------------------------------------------------------
% ----------------------    Appendix A   -------------------------------------
%------------------------------------------------------------------------------

\section{Posterior Analysis Using MCMC}\label{app:full}
%-----------------------------------------------------------------------------

The joint posterior density for the binary quantile model with 
misclassification, obtained by combining the complete data likelihood with the 
prior distributions on the parameter vector $\Theta=(\beta, 
\delta_{01},\delta_{10})$, is given by,
%--------------------------
\begin{eqnarray}
%\begin{split}
	&& \pi(\Theta, z,y,w|y^{obs}) 
	= \prod_{i=1}^{n} \bigg\{ 
	f(y_{i}^{obs}|y_{i},\delta_{01},\delta_{10}) f(y_{i}|z_{i})  
	f_{N}(z_{i}|\beta,w_{i}) f_{\mathcal{E}}(w_{i})\bigg\} 
	 \, \pi_{N}(\beta|\beta_{0},B_{0}) \nonumber \\
	&& \hspace{1.4in} \times \; \pi_{B}(\delta_{01}|\kappa_{1},\kappa_{2})  
	 \,	 \pi_{B}(\delta_{10}|\kappa_{3},\kappa_{4})\nonumber \\
	&& \propto \prod_{i=1}^{n} \bigg\{   
	\big[ 	\delta_{01}^{y_{i}(1-y^{obs}_{i})} 
	(1-\delta_{01})^{y_{i}^{obs} y_{i}}   \times 
	\delta_{10}^{y^{obs}_{i}(1-y_{i})} 
	(1-\delta_{10})^{(1-y_{i}^{obs})(1-y_{i})} \big] \nonumber \\
	&& \quad \times \, \big[ I(z_{i}>0)I(y_{i}=1) + I(z_{i} \leq 
	0) I(y_{i}=0) \big] 
	\, N(z_{i}|x'_{i}\beta + \theta w_{i}, \tau^{2} w_{i}) \;
	\mathcal{E}(w_{i}|1) \bigg\}  \nonumber \\
	&& \quad \times \, \pi_{N}(\beta|\beta_{0},B_{0})
	\, \pi_{B}(\delta_{01}|\kappa_{1},\kappa_{2}) 
	\, \pi_{B}(\delta_{10}|\kappa_{3},\kappa_{4}),
%\end{split}
\label{eq:bqrMis-AugJP} 
\end{eqnarray}
%--------------------------
where $\pi_{N}(\cdot)$ and $\pi_{B}(\cdot)$ denote the prior densities for 
normal and Beta distributions, respectively.

To derive the conditional posterior distributions from the joint 
density~\eqref{eq:bqrMis-AugJP}, we isolate all terms involving the parameter 
of interest, while treating the remaining unknown quantities as fixed, and 
then identify the resulting distribution corresponding to that parameter. We 
follow this approach and derive the conditional posteriors as outlined in 
Algorithm~\ref{alg:algorithm2}.

We begin by deriving the full conditional distribution of $\beta$, denoted  
$\pi(\beta|z,w)$. This density is proportional to the product $f(z|\beta,w) 
\, \pi(\beta)$, and its kernel can be written as follows: 
%------------------------------
\begin{eqnarray*}
	\pi(\beta|z,w)
	& \propto &
	\exp\bigg[-\frac{1}{2} \bigg\{ \sum_{i=1}^{n}  \bigg( \frac{z_{i} - 
	x'_{i}\beta - \theta w_{i}}
	{\tau \sqrt{w_{i}}} \bigg)^2  +
	(\beta - \beta_{0} )' B_{0}^{-1} (\beta - \beta_{0} )   
	\bigg\}  \bigg]\\
	& \propto &
	\exp\bigg[-\frac{1}{2} \bigg\{ \beta'
	\bigg( \sum_{i=1}^{n} \frac{x_{i} x'_{i}}{\tau^{2} w_{i}} + 
	B_{0}^{-1}\bigg) \beta
	- \beta' \bigg( \sum_{i=1}^{n} \frac{x_{i}(z_{i}-\theta 
	w_{i})}{\tau^{2} w_{i}}
	+ B_{0}^{-1} \beta_{0}  \bigg)     \\
	& & \qquad - \bigg( \sum_{i=1}^{n} \frac{x'_{i}(z_{i}-\theta 
	w_{i})}{\tau^{2} w_{i}}
	+ \beta'_{0}B_{0}^{-1}  \bigg) \beta    \bigg\}          \bigg]\\
	& \propto &
	\exp\bigg[ -\frac{1}{2} \bigg\{ \beta'\tilde{B}^{-1}\beta
	- \beta'\tilde{B}^{-1}\tilde{\beta} - \tilde{\beta}'
	\tilde{B}^{-1}\beta
	\bigg\}  \bigg], \\
	& \propto &   \exp\bigg[ -\frac{1}{2}
	\bigg\{ \beta'\tilde{B}^{-1}\beta
	- \beta'\tilde{B}^{-1}\tilde{\beta}
	- \tilde{\beta}' \tilde{B}^{-1}\beta
	+ \tilde{\beta}'\tilde{B}^{-1} \tilde{\beta}
	- \tilde{\beta}'\tilde{B}^{-1} \tilde{\beta}
	\bigg\}  \bigg]\\
	& \propto &
	\exp\bigg[ -\frac{1}{2}
	\bigg\{ (\beta - \tilde{\beta})'
	\tilde{B}^{-1}(\beta - \tilde{\beta})
	\bigg\}  \bigg],
\end{eqnarray*}
%------------------------------
where, in the second line we omit all terms not involving $\beta$, and in 
the third line we introduce two terms: $\tilde{B}^{-1} = \big(\sum_{i=1}^{n} 
\frac{x_{i} x'_{i}}{\tau^{2} w_{i}} + B_{0}^{-1} \big)$ and $\tilde{\beta} = 
\tilde{B} \big( \sum_{i=1}^{n} \frac{x_{i}(z_{i} - \theta w_{i})}{\tau^{2} 
w_{i}} + B_{0}^{-1} \beta_{0} \big)$. In the fourth line, we add and subtract 
$\tilde{\beta}'\tilde{B}^{-1}\tilde{\beta}$, and lastly in the fifth line we 
complete the square. The result is kernel of a Gaussian or normal density and 
hence $\beta|z,w\sim N(\tilde{\beta}, \tilde{B})$.

%-------------------------------------------------------------------------------

The full conditional density for the false negative rate $\delta_{01}$, given 
by $\pi(\delta_{01}|y,y^{obs})$ is proportional to 
$f(y^{obs}|y,\delta_{01}) \pi(\delta_{01})$. The kernel can be written as,
%------------------------------
\begin{eqnarray*}
\pi(\delta_{01}|\beta,y,y^{obs}) & \propto & \prod_{i=1}^{n} 
\big[ 	\delta_{01}^{y_{i}(1-y^{obs}_{i})} 
(1-\delta_{01})^{y_{i}^{obs} y_{i}}  \big] \times 
\delta_{01}^{(\kappa_{1}-1)} (1 - \delta_{01})^{(\kappa_{2}-1)} \\
& \propto & \delta_{01}^{ 
 \sum_{i=1}^{n} (y_{i}(1-y^{obs}_{i})) + \kappa_{1}-1 }\;
(1-\delta_{01})^{\sum_{i=1}^{n} (y_{i}y^{obs}_{i}) + \kappa_{2}-1} \\
& \propto & \delta_{01}^{\tilde{\kappa}_{1}-1} 
(1-\delta_{01})^{\tilde{\kappa}_{2}-1},
\end{eqnarray*}
%------------------------------
where $\tilde{\kappa}_{1} = \sum_{i=1}^{n} (y_{i}(1-y^{obs}_{i})) + 
\kappa_{1}$, and $\tilde{\kappa}_{2} = \sum_{i=1}^{n} (y_{i}y^{obs}_{i}) + 
\kappa_{2}$. The last expression is recognized as the kernel of a Beta 
distribution, hence, $\delta_{01}|y,y^{obs} \sim B(\tilde{\kappa}_{1}, 
\tilde{\kappa}_{2})$.
%-------------------------------------------------------------------------------

Similarly, the full conditional density for the false positive rate 
$\delta_{10}$, given by $\pi(\delta_{10}|y,y^{obs})$ is proportional to 
$f(y^{obs}|y,\delta_{10}) \pi(\delta_{10})$. The kernel can be written as,
%------------------------------
\begin{eqnarray*}
	\pi(\delta_{10}|\beta,y,y^{obs}) & \propto & \prod_{i=1}^{n} 
	\big[ 	\delta_{10}^{y_{i}^{obs}(1-y_{i})} 
	(1-\delta_{10})^{(1 - y_{i}^{obs})(1 - y_{i})}  \big] \times 
	\delta_{10}^{(\kappa_{3}-1)} (1 - \delta_{10})^{(\kappa_{4}-1)} \\
	& \propto & \delta_{10}^{ 
		\sum_{i=1}^{n} (y^{obs}_{i}(1-y_{i})) + \kappa_{3}}-1 \;
	(1-\delta_{10})^{\sum_{i=1}^{n} (1-y_{i}) (1 -y^{obs}_{i}) + 
	\kappa_{4}-1} \\
	& \propto & \delta_{10}^{\tilde{\kappa}_{3}-1} 
	(1-\delta_{10})^{\tilde{\kappa}_{4}-1},
\end{eqnarray*}
%------------------------------
where $\tilde{\kappa}_{3} = \sum_{i=1}^{n} (y^{obs}_{i}(1-y_{i})) + 
\kappa_{3}$, and $\tilde{\kappa}_{4} = \sum_{i=1}^{n} (1-y_{i}) 
(1-y^{obs}_{i}) + \kappa_{4}$. The last expression is recognized as the 
kernel of a Beta distribution, hence, $\delta_{10}|y,y^{obs} \sim 
B(\tilde{\kappa}_{3}, \tilde{\kappa}_{4})$.
%-------------------------------------------------------------------------------

To sample the unobserved true binary variable $y$ from a tractable 
conditional distribution, we marginalize the conditional density with  
respect to $(z,w)$; thus collapsing the Gibbs sampler. Besides, $y$ cannot be sampled jointly, so each $y_{i}$ 
must be sampled individually. Extracting terms involving $y_{i}$ from the 
joint posterior~\eqref{eq:bqrMis-AugJP}, the conditional density for $y_{i}$ 
can be written as, 
%------------------------------
\begin{eqnarray*}
\pi(y_{i}|\Theta,y^{obs}_{i}) &\propto & f(y^{obs}_{i}|y_{i}, \delta_{01}, 
\delta_{10}) f_{AL}(y_{i}|\beta)\\
& \propto & 
\big[ 	\delta_{01}^{y_{i}(1-y^{obs}_{i})} 
(1-\delta_{01})^{y_{i}^{obs} y_{i}}   \times 
\delta_{10}^{y^{obs}_{i}(1-y_{i})} 
(1-\delta_{10})^{(1-y_{i}^{obs})(1-y_{i})} \big] \nonumber \\
&& \times \, \big[ 1 - F_{AL}(-x'_{i}\beta)\big]^{y_{i}} 
\big[ F_{AL}(-x'_{i}\beta)\big]^{(1 - y_{i})}.
\end{eqnarray*}
%------------------------------
Based on the above equation we can see that,
%------------------------------
\begin{eqnarray*}
\pi(y_{i}=1|\Theta,y^{obs}_{i}) & \propto & \delta_{01}^{(1-y^{obs}_{i})} 
(1-\delta_{01})^{y^{obs}_{i}} \times  \big[ 1 - F_{AL}(-x'_{i}\beta)\big], 
\quad \mathrm{and}\\
\pi(y_{i}=0|\Theta,y^{obs}_{i}) & \propto & \delta_{10}^{y^{obs}_{i}} 
(1-\delta_{10})^{(1-y^{obs}_{i})} \times  \big[ F_{AL}(-x'_{i}\beta)\big].
\end{eqnarray*}
%------------------------------	
Since $y_{i} \in \{0,1\}$, the probabilities must sum to 1. Therefore, 
dividing $\pi(y_{i}=1|\Theta,y^{obs}_{i})$ by the sum of probabilities, we 
have,
%------------------------------
\begin{eqnarray*}
	\pi(y_{i}=1|\Theta,y^{obs}_{i}) & = &\frac{\delta_{01}^{(1-y^{obs}_{i})} 
		(1-\delta_{01})^{y^{obs}_{i}} \times  
		\psi_{i}}{\delta_{01}^{(1-y^{obs}_{i})} 
		(1-\delta_{01})^{y^{obs}_{i}} \times  \psi_{i} + 
		\delta_{10}^{y^{obs}_{i}} 
		(1-\delta_{10})^{(1-y^{obs}_{i})} \times  \big[ 
		1-\psi_{i}\big] } = \Psi_{i}	
\end{eqnarray*}
%------------------------------	
where $\psi_{i} = 1 - F_{AL}(-x'_{i}\beta)$. Therefore, 
$y_{i}|\Theta,y^{obs}_{i}$ can be sampled from a Bernoulli distribution with 
success probability $\Psi_{i}$. Hence, $y_{i}|\Theta,y^{obs}_{i} \sim 
\mathrm{Bernoulli}(\Psi_{i})$ for $i=1,2,\cdots,n$. 

%-------------------------------------------------------------------------------

The full conditional distribution of $w$,
denoted by $\pi(w|z,\beta_{p})$ is proportional to $f(z|\beta,w) \pi(w)$. 
However, the sampling of $w$ must be performed element-wise.
The kernel for each $w_{i}$ can be derived as follows,
%------------------------------
\begin{eqnarray*}
	\pi(w_{i}|z,\beta)
	& \propto &
	w_{i}^{-1/2} \exp\bigg[ -\frac{1}{2} \bigg(\frac{z_{i} - x'_{i}\beta 
	- \theta w_{i}}
	{\tau \sqrt{w_{i}}}   \bigg)^{2}  - w_{i}     \bigg]\\
	& \propto &
	w_{i}^{-1/2} \exp\bigg[ -\frac{1}{2} \bigg( \frac{ (z_{i} - 
	x'_{i}\beta)^{2} + \theta^{2} w_{i}^{2}
		- 2\theta w_{i} (z_{i} - x'_{i}\beta) }{\tau^{2} w_{i}} + 2 
		w_{i}  \bigg)    \bigg]\\
	& \propto &
	w_{i}^{-1/2} \exp\bigg[ -\frac{1}{2}  \bigg\{
	\frac{ (z_{i} - x'_{i}\beta)^{2}}{\tau^{2}} \; w_{i}^{-1}
	+ \bigg( \frac{\theta^{2}}{\tau^{2}} + 2 \bigg)
	w_{i} \bigg\}    \bigg]\\
	& \propto &
	w_{i}^{-1/2} \exp\bigg[ -\frac{1}{2} \big\{ \tilde{\lambda_{i}} \, 
	w_{i}^{-1} +
	\tilde{\eta} \, w_{i} \big\}
	\bigg].
\end{eqnarray*}
%------------------------------
The last expression can be recognized as the kernel of the GIG distribution,
where, $\tilde{\lambda_{i}} = \frac{ ( z_{i} - 	x'_{i}\beta )^2 }{\tau^{2}}$ 
and $\tilde{\eta} = \Big( \frac{\theta^2}{\tau^{2}} + 2 \Big)$. Hence, we 
have $w_{i}|z,\beta \sim GIG(0.5, \tilde{\lambda_{i}}, \tilde{\eta})$.

%-------------------------------------------------------------------------------

Finally, we need to find the full conditional distribution of continuous 
latent variable $z$. As with $y$, each elements of $z$ has to be sampled 
individually. Collecting terms involving $z_{i}$ from the joint 
posterior~\eqref{eq:bqrMis-AugJP}, we obtain,
%--------------------------
\begin{eqnarray*}
	%\begin{split}
	\pi(z_{i}|\beta,w_{i},y_{i}) 
	& \propto & f(y_{i}|z_{i})  
	f_{N}(z_{i}|\beta,w_{i})  \\
	& \propto & \big[ I(z_{i}>0)I(y_{i}=1) + I(z_{i} \leq 
	0) I(y_{i}=0) \big] 
	\, N(z_{i}|x'_{i}\beta + \theta w_{i}, \tau^{2} w_{i})  \\
	&\propto & \big[ I(z_{i}>0)I(y_{i}=1) + I(z_{i} \leq 
	0) I(y_{i}=0) \big] \exp\bigg[-\frac{1}{2} \bigg\{ \sum_{i=1}^{n}  \bigg( 
	\frac{z_{i} - x'_{i}\beta - \theta w_{i}}
	{\tau \sqrt{w_{i}}} \bigg)^2 \bigg\}  \bigg].
\end{eqnarray*}
%--------------------------
The last expression immediately implies that,
%--------------------------
\begin{eqnarray*}
	%\begin{split}
	\pi(z_{i}|\beta,w_{i},y_{i}=0) 
	&\propto &  \exp\bigg[-\frac{1}{2} \bigg\{ \sum_{i=1}^{n}  \bigg( 
	\frac{z_{i} - x'_{i}\beta - \theta w_{i}}
	{\tau \sqrt{w_{i}}} \bigg)^2 \bigg\}  \bigg] \; I(z_{i} \le 0) \\
	\pi(z_{i}|\beta,w_{i},y_{i}=1) 
	&\propto &  \exp\bigg[-\frac{1}{2} \bigg\{ \sum_{i=1}^{n}  \bigg( 
	\frac{z_{i} - x'_{i}\beta - \theta w_{i}}
	{\tau \sqrt{w_{i}}} \bigg)^2 \bigg\}  \bigg] \; I(z_{i} > 0).
\end{eqnarray*}
%--------------------------
These are kernels of normal distributions truncated-from-above at zero and 
truncated-from-below at zero. Hence, we can write, 
%--------------------------
\begin{eqnarray*}
	z_{i}|\beta,w_{i},y_{i}=0
	& \sim  &  TN(x'_{i}\beta + \theta w_{i}, \tau^{2}w_{i}, -\infty,0) \\
	z_{i}|\beta,w_{i},y_{i}=1 
	& \sim &  TN(x'_{i}\beta + \theta w_{i}, \tau^{2}w_{i}, 0, \infty),
\end{eqnarray*}
%--------------------------
where $TN(\mu,\sigma^{2},L,U)$ denotes a truncated normal distribution with 
mean $\mu$, variance $\sigma^{2}$, lower truncation bound $L$, and upper 
truncation bound $U$.

\clearpage
\section{MCMC Trace Plots for Simulation Studies}
\label{app:trace}
\begin{figure}[ht]
\centering
\includegraphics[angle=270,width=1\textwidth]{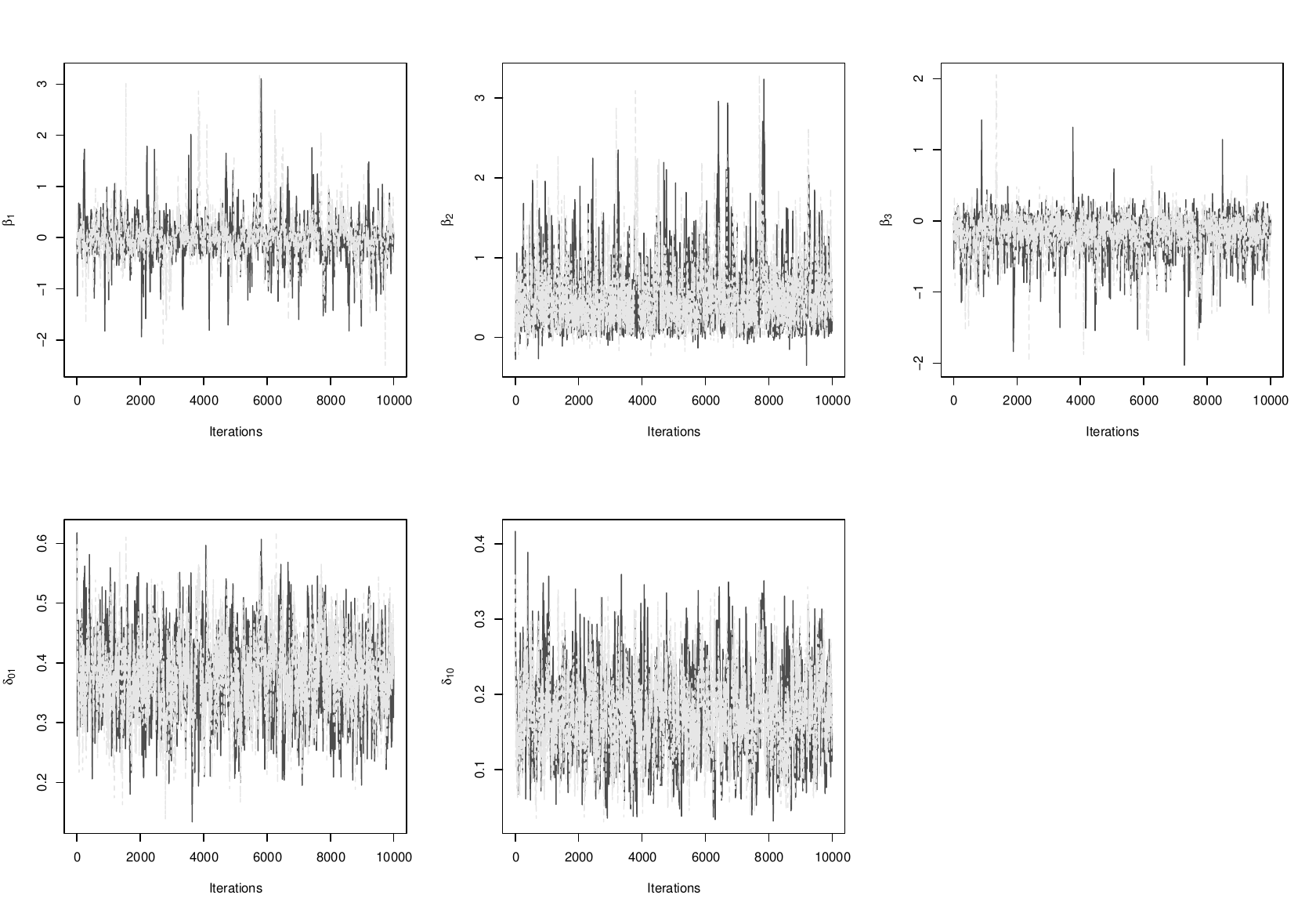}
\caption{Representative trace plots of the parameters in the misclassification model, based on two MCMC chains at the 50-th quantile, when $n_{pess}=30$, $(\delta_{01}=0.4,\delta_{10}=0.2)$, and $Var(\beta)= 10 I_{k}$. The first row presents the trace plots of the regression coefficients, while the second row shows the trace plots of the misclassification parameters.}
\label{mcmc}
\end{figure}

%------------------------------ Supplementary Material ---------------------------------

\newpage
\section{Supplementary Simulation Results}
\label{app:sim}

%-------------------------------------------------------------------------------
\begin{table}[ht]
\centering
\caption{Average biases of posterior estimates of regression coefficients from the two models when $\delta_{01}=0.4$ and $\delta_{10}=0.2$ under different settings of prior variance and prior effective sample size ($n_{\text{pess}}$).} 
\label{Table:Case1_AB}
\begin{tabular}{ccc ccc ccc}
\toprule
 &  &  & \multicolumn{3}{c}{BQR} & \multicolumn{3}{c}{BQRM} \\
\cmidrule(lr){4-6} \cmidrule(lr){7-9}
%  \midrule
  $n_{\text{pess}}$ & $\sigma^{2}_{\beta}$ & $p$ & $\beta_{1}$ & $\beta_{2}$ & $\beta_{3}$ & $\beta_{1}$ & $\beta_{2}$ & $\beta_{3}$ \\ 
  \midrule
30 & 5 & 0.10 & -12.84 & -0.66 & 0.32 & -9.64 & -0.67 & 0.26 \\ 
   &  & 0.25 & -2.17 & -0.66 & 0.33 & -1.57 & -0.10 & 0.03 \\ 
   &  & 0.50 & -0.45 & -0.68 & 0.33 & -0.02 & -0.35 & 0.16 \\ 
   &  & 0.75 & 0.68 & -0.68 & 0.33 & 1.53 & -0.08 & 0.01 \\ 
   &  & 0.90 & 7.44 & -0.55 & 0.27 & 9.98 & -0.48 & 0.24 \\ 
     \midrule
%30 & 10 & 0.10 & -15.09 & -0.56 & 0.25 & -12.27 & -0.49 & 0.24 \\ 
%   &  & 0.25 & -2.49 & -0.63 & 0.29 & -1.93 & 0.02 & -0.06 \\ 
%   &  & 0.50 & -0.45 & -0.69 & 0.36 & -0.03 & -0.39 & 0.21 \\ 
 %  &  & 0.75 & 0.66 & -0.66 & 0.32 & 1.64 & 0.06 & -0.06 \\ 
 %  &  & 0.90 & 8.73 & -0.49 & 0.23 & 13.22 & -0.43 & 0.24 \\ 
 %    \midrule
50 & 5 & 0.10 & -17.41 & -0.56 & 0.26 & -11.51 & -0.59 & 0.31 \\ 
   &  & 0.25 & -2.47 & -0.65 & 0.31 & -1.78 & -0.07 & 0.03 \\ 
   &  & 0.50 & -0.43 & -0.68 & 0.34 & -0.00 & -0.34 & 0.17 \\ 
   &  & 0.75 & 0.68 & -0.65 & 0.32 & 1.48 & 0.05 & -0.05 \\ 
   &  & 0.90 & 8.25 & -0.50 & 0.23 & 10.60 & -0.45 & 0.20 \\ 
   \midrule
50 & 10 & 0.10 & -17.23 & -0.57 & 0.27 & -13.10 & -0.51 & 0.26 \\ 
   &  & 0.25 & -2.72 & -0.60 & 0.33 & -2.06 & 0.04 & 0.09 \\ 
   &  & 0.50 & -0.44 & -0.70 & 0.35 & -0.02 & -0.38 & 0.19 \\ 
   &  & 0.75 & 0.63 & -0.64 & 0.34 & 1.59 & 0.17 & 0.00 \\ 
   &  & 0.90 & 8.68 & -0.50 & 0.21 & 13.77 & -0.46 & 0.20 \\ 
    \midrule
100 & 5 & 0.10 & -15.65 & -0.55 & 0.29 & -10.72 & -0.54 & 0.26 \\ 
   &  & 0.25 & -2.40 & -0.64 & 0.34 & -1.79 & -0.07 & 0.09 \\ 
   &  & 0.50 & -0.45 & -0.67 & 0.34 & -0.01 & -0.33 & 0.18 \\ 
   &  & 0.75 & 0.61 & -0.67 & 0.34 & 1.40 & 0.05 & -0.00 \\ 
   &  & 0.90 & 7.22 & -0.54 & 0.31 & 9.78 & -0.37 & 0.19 \\ 
    \midrule
100 & 10 & 0.10 & -18.17 & -0.55 & 0.27 & -13.59 & -0.44 & 0.26 \\ 
   &  & 0.25 & -2.27 & -0.63 & 0.32 & -1.79 & 0.04 & -0.04 \\ 
   &  & 0.50 & -0.45 & -0.69 & 0.36 & -0.03 & -0.35 & 0.21 \\ 
   &  & 0.75 & 0.63 & -0.67 & 0.32 & 1.39 & 0.07 & -0.09 \\ 
   &  & 0.90 & 7.55 & -0.57 & 0.32 & 12.18 & -0.31 & 0.24 \\ 
   \bottomrule
\end{tabular}
\end{table}
%-------------------------------------------------------------------------------

%-------------------------------------------------------------------------------
%\beginsupplement
\begin{table}[ht]
\centering
\caption{Mean squared errors (MSE) of posterior estimates of regression coefficients from the two models when $\delta_{01}=0.4$ and $\delta_{10}=0.2$ under different settings of prior variance and prior effective sample size ($n_{\text{pess}}$).} 
\label{Table:Case1_MSE}
\begin{tabular}{ccc ccc ccc}
\toprule
 &  &  & \multicolumn{3}{c}{BQR} & \multicolumn{3}{c}{BQRM} \\
\cmidrule(lr){4-6} \cmidrule(lr){7-9}
%  \midrule
$n_{\text{pess}}$ & $\sigma^{2}_{\beta}$ & $p$ & $\beta_{1}$ & $\beta_{2}$ & $\beta_{3}$ & $\beta_{1}$ & $\beta_{2}$ & $\beta_{3}$ \\ 
%ESS & $\sigma^{2}_{\beta}$ & p & Naive.B0 & Naive.B1 & Naive.B2 & Mis.B0 & Mis.B1 & Mis.B2 \\ 
  \midrule
30 & 5 & 0.10 & 332.11 & 0.60 & 0.25 & 122.17 & 0.56 & 0.18 \\ 
   &  & 0.25 & 7.84 & 0.48 & 0.13 & 5.46 & 0.12 & 0.11 \\ 
   &  & 0.50 & 0.21 & 0.47 & 0.11 & 0.03 & 0.16 & 0.05 \\ 
   &  & 0.75 & 0.76 & 0.49 & 0.12 & 4.58 & 0.14 & 0.07 \\ 
   &  & 0.90 & 78.38 & 0.42 & 0.14 & 118.90 & 0.36 & 0.21 \\ 
    \midrule
%30 & 10 & 0.10 & 421.98 & 0.55 & 0.19 & 186.67 & 0.48 & 0.26 \\ 
 %  &  & 0.25 & 9.84 & 0.45 & 0.11 & 8.09 & 0.17 & 0.09 \\ 
 %  &  & 0.50 & 0.21 & 0.49 & 0.13 & 0.02 & 0.19 & 0.07 \\ 
 %  &  & 0.75 & 0.72 & 0.46 & 0.11 & 4.83 & 0.18 & 0.11 \\ 
 %  &  & 0.90 & 95.92 & 0.35 & 0.14 & 197.07 & 0.43 & 0.21 \\ 
 %   \midrule
50 & 5 & 0.10 & 472.24 & 0.52 & 0.20 & 161.38 & 0.49 & 0.21 \\ 
   &  & 0.25 & 9.39 & 0.46 & 0.12 & 6.35 & 0.11 & 0.10 \\ 
   &  & 0.50 & 0.20 & 0.47 & 0.12 & 0.02 & 0.15 & 0.06 \\ 
   &  & 0.75 & 0.71 & 0.44 & 0.11 & 3.70 & 0.15 & 0.09 \\ 
   &  & 0.90 & 92.53 & 0.40 & 0.16 & 130.91 & 0.36 & 0.20 \\ 
    \midrule
50 & 10 & 0.10 & 488.01 & 0.53 & 0.18 & 210.28 & 0.52 & 0.21 \\ 
   &  & 0.25 & 10.75 & 0.41 & 0.14 & 7.69 & 0.17 & 0.13 \\ 
   &  & 0.50 & 0.20 & 0.49 & 0.13 & 0.02 & 0.17 & 0.06 \\ 
   &  & 0.75 & 0.65 & 0.44 & 0.13 & 4.90 & 0.19 & 0.11 \\ 
   &  & 0.90 & 92.67 & 0.37 & 0.14 & 210.40 & 0.43 & 0.23 \\ 
   \midrule
   100 & 5 & 0.10 & 422.55 & 0.50 & 0.21 & 154.18 & 0.45 & 0.22 \\ 
   &  & 0.25 & 9.05 & 0.45 & 0.14 & 6.88 & 0.10 & 0.10 \\ 
   &  & 0.50 & 0.21 & 0.46 & 0.12 & 0.03 & 0.16 & 0.06 \\ 
   &  & 0.75 & 0.63 & 0.47 & 0.13 & 3.90 & 0.14 & 0.10 \\ 
   &  & 0.90 & 79.12 & 0.45 & 0.15 & 122.84 & 0.31 & 0.19 \\
    \midrule
100 & 10 & 0.10 & 524.01 & 0.52 & 0.19 & 236.08 & 0.51 & 0.28 \\ 
   &  & 0.25 & 8.07 & 0.46 & 0.12 & 6.53 & 0.20 & 0.11 \\ 
   &  & 0.50 & 0.21 & 0.48 & 0.14 & 0.02 & 0.16 & 0.07 \\ 
   &  & 0.75 & 0.69 & 0.47 & 0.12 & 4.04 & 0.18 & 0.13 \\ 
   &  & 0.90 & 81.72 & 0.45 & 0.16 & 182.37 & 0.38 & 0.37 \\ 
   \bottomrule
\end{tabular}
\end{table}
%-------------------------------------------------------------------------------

%-------------------------------------------------------------------------------
\begin{table}[ht]
\centering
\caption{Coverage rates of posterior estimates of regression coefficients from the two models when $\delta_{01}=0.4$ and $\delta_{10}=0.2$ under different settings of prior variance and prior effective sample size ($n_{\text{pess}}$).} 
\label{Table:Case1_Coverage}
\begin{tabular}{ccc ccc ccc}
\toprule
 &  &  & \multicolumn{3}{c}{BQR} & \multicolumn{3}{c}{BQRM} \\
\cmidrule(lr){4-6} \cmidrule(lr){7-9}
%  \midrule
  $n_{\text{pess}}$ & $\sigma^{2}_{\beta}$ & $p$ & $\beta_{1}$ & $\beta_{2}$ & $\beta_{3}$ & $\beta_{1}$ & $\beta_{2}$ & $\beta_{3}$ \\ 
  \midrule
30 & 5 & 0.10 & 0.00 & 0.41 & 0.43 & 1.00 & 1.00 & 1.00 \\ 
   &  & 0.25 & 0.00 & 0.32 & 0.41 & 1.00 & 1.00 & 1.00 \\ 
   &  & 0.50 & 0.00 & 0.00 & 0.26 & 1.00 & 1.00 & 0.99 \\ 
   &  & 0.75 & 0.98 & 0.30 & 0.50 & 1.00 & 1.00 & 1.00 \\ 
   &  & 0.90 & 0.30 & 0.67 & 0.66 & 1.00 & 1.00 & 1.00 \\ 
     \midrule
%30 & 10 & 0.10 & 0.00 & 0.44 & 0.51 & 1.00 & 1.00 & 1.00 \\ 
%   &  & 0.25 & 0.00 & 0.29 & 0.45 & 1.00 & 1.00 & 1.00 \\ 
%   &  & 0.50 & 0.00 & 0.00 & 0.15 & 1.00 & 0.99 & 0.96 \\ 
%   &  & 0.75 & 0.96 & 0.33 & 0.50 & 1.00 & 1.00 & 1.00 \\ 
%   &  & 0.90 & 0.19 & 0.73 & 0.74 & 1.00 & 1.00 & 1.00 \\ 
%     \midrule
50 & 5 & 0.10 & 0.00 & 0.54 & 0.58 & 1.00 & 1.00 & 1.00 \\ 
   &  & 0.25 & 0.00 & 0.31 & 0.46 & 1.00 & 1.00 & 1.00 \\ 
   &  & 0.50 & 0.00 & 0.00 & 0.18 & 1.00 & 1.00 & 0.98 \\ 
   &  & 0.75 & 1.00 & 0.34 & 0.54 & 1.00 & 1.00 & 1.00 \\ 
   &  & 0.90 & 0.25 & 0.65 & 0.65 & 1.00 & 1.00 & 1.00 \\ 
    \midrule
50 & 10 & 0.10 & 0.00 & 0.50 & 0.57 & 1.00 & 1.00 & 1.00 \\ 
   &  & 0.25 & 0.00 & 0.35 & 0.47 & 1.00 & 1.00 & 0.99 \\ 
   &  & 0.50 & 0.00 & 0.00 & 0.10 & 1.00 & 1.00 & 0.99 \\ 
   &  & 0.75 & 0.99 & 0.41 & 0.44 & 1.00 & 1.00 & 0.99 \\ 
   &  & 0.90 & 0.17 & 0.70 & 0.74 & 1.00 & 1.00 & 1.00 \\ 
    \midrule
100 & 5 & 0.10 & 0.00 & 0.50 & 0.52 & 1.00 & 1.00 & 1.00 \\ 
   &  & 0.25 & 0.00 & 0.31 & 0.36 & 1.00 & 1.00 & 1.00 \\ 
   &  & 0.50 & 0.00 & 0.01 & 0.17 & 1.00 & 0.99 & 0.98 \\ 
   &  & 0.75 & 1.00 & 0.34 & 0.42 & 1.00 & 1.00 & 1.00 \\ 
   &  & 0.90 & 0.33 & 0.57 & 0.60 & 1.00 & 1.00 & 1.00 \\ 
   \midrule
100 & 10 & 0.10 & 0.00 & 0.55 & 0.58 & 1.00 & 1.00 & 1.00 \\ 
   &  & 0.25 & 0.00 & 0.30 & 0.43 & 1.00 & 1.00 & 1.00 \\ 
   &  & 0.50 & 0.00 & 0.00 & 0.09 & 1.00 & 1.00 & 1.00 \\ 
   &  & 0.75 & 1.00 & 0.30 & 0.49 & 1.00 & 1.00 & 1.00 \\ 
   &  & 0.90 & 0.29 & 0.61 & 0.63 & 1.00 & 1.00 & 1.00 \\
   \bottomrule
\end{tabular}
\end{table}
%-------------------------------------------------------------------------------

%-------------------------------------------------------------------------------
%                       Second
%-------------------------------------------------------------------------------

%-------------------------------------------------------------------------------
\begin{table}[ht]
\centering
\caption{Average biases of posterior estimates of regression coefficients from the two models when $\delta_{01}=0.3$ and $\delta_{10}=0.2$ under different settings of prior variance and prior effective sample size ($n_{\text{pess}}$).} 
\label{Table:Case2_AB}
  \begin{tabular}{ccc ccc ccc}
\toprule
 &  &  & \multicolumn{3}{c}{BQR} & \multicolumn{3}{c}{BQRM} \\
\cmidrule(lr){4-6} \cmidrule(lr){7-9}
%  \midrule
  $n_{\text{pess}}$ & $\sigma^{2}_{\beta}$ & $p$ & $\beta_{1}$ & $\beta_{2}$ & $\beta_{3}$ & $\beta_{1}$ & $\beta_{2}$ & $\beta_{3}$ \\ 
  \midrule
30 & 5 & 0.10 & -12.24 & -0.51 & 0.26 & -10.78 & -0.41 & 0.17 \\ 
   &  & 0.25 & -1.48 & -0.60 & 0.29 & -1.45 & 0.13 & -0.09 \\ 
   &  & 0.50 & -0.21 & -0.63 & 0.32 & -0.01 & -0.31 & 0.16 \\ 
   &  & 0.75 & 0.83 & -0.59 & 0.31 & 1.53 & 0.22 & -0.07 \\ 
   &  & 0.90 & 8.40 & -0.42 & 0.20 & 10.95 & -0.31 & 0.14 \\ 
     \midrule
%30 & 10 & 0.10 & -12.38 & -0.54 & 0.27 & -12.77 & -0.39 & 0.19 \\ 
%%   &  & 0.25 & -1.45 & -0.62 & 0.31 & -1.46 & 0.22 & -0.12 \\ 
  % &  & 0.50 & -0.23 & -0.63 & 0.32 & -0.02 & -0.26 & 0.13 \\ 
  % &  & 0.75 & 0.92 & -0.56 & 0.27 & 1.98 & 0.38 & -0.20 \\ 
  % &  & 0.90 & 8.62 & -0.39 & 0.25 & 12.96 & -0.21 & 0.13 \\
  %  \midrule
50 & 5 & 0.10 & -11.77 & -0.49 & 0.27 & -10.41 & -0.40 & 0.21 \\ 
   &  & 0.25 & -1.78 & -0.57 & 0.28 & -1.89 & 0.15 & -0.06 \\ 
   &  & 0.50 & -0.22 & -0.62 & 0.30 & -0.01 & -0.23 & 0.11 \\ 
   &  & 0.75 & 0.91 & -0.56 & 0.28 & 1.77 & 0.25 & -0.10 \\ 
   &  & 0.90 & 8.45 & -0.47 & 0.19 & 10.87 & -0.27 & 0.12 \\
    \midrule
50 & 10 & 0.10 & -11.27 & -0.54 & 0.24 & -11.90 & -0.24 & 0.16 \\ 
   &  & 0.25 & -1.50 & -0.59 & 0.29 & -1.64 & 0.27 & -0.13 \\ 
   &  & 0.50 & -0.22 & -0.62 & 0.31 & -0.01 & -0.23 & 0.11 \\ 
   &  & 0.75 & 0.90 & -0.57 & 0.28 & 1.83 & 0.33 & -0.18 \\ 
   &  & 0.90 & 8.66 & -0.42 & 0.24 & 12.85 & -0.17 & 0.12 \\
   \midrule
100 & 5 & 0.10 & -13.67 & -0.39 & 0.21 & -12.05 & -0.28 & 0.24 \\ 
   &  & 0.25 & -1.81 & -0.55 & 0.30 & -1.90 & 0.18 & -0.08 \\ 
   &  & 0.50 & -0.21 & -0.64 & 0.33 & -0.01 & -0.28 & 0.16 \\ 
   &  & 0.75 & 0.89 & -0.57 & 0.30 & 1.73 & 0.25 & -0.09 \\ 
   &  & 0.90 & 8.19 & -0.38 & 0.24 & 11.02 & -0.13 & 0.11 \\
   \midrule
100 & 10 & 0.10 & -13.85 & -0.44 & 0.19 & -14.25 & -0.27 & 0.02 \\ 
   &  & 0.25 & -1.68 & -0.58 & 0.30 & -1.83 & 0.18 & -0.07 \\ 
   &  & 0.50 & -0.22 & -0.61 & 0.30 & 0.01 & -0.23 & 0.10 \\ 
   &  & 0.75 & 0.93 & -0.57 & 0.30 & 1.69 & 0.27 & -0.10 \\ 
   &  & 0.90 & 8.12 & -0.43 & 0.23 & 12.78 & -0.06 & 0.13 \\
%   \midrule
\bottomrule
\end{tabular}
\end{table}
%-------------------------------------------------------------------------------

%-------------------------------------------------------------------------------
\begin{table}[ht]
\centering
\caption{Mean squared errors (MSE) of posterior estimates of regression coefficients from the two models when $\delta_{01}=0.3$ and $\delta_{10}=0.2$ under different settings of prior variance and prior effective sample size ($n_{\text{pess}}$).} 
\label{Table:Case2_MSE}
\begin{tabular}{ccc ccc ccc}
\toprule
 &  &  & \multicolumn{3}{c}{BQR} & \multicolumn{3}{c}{BQRM} \\
\cmidrule(lr){4-6} \cmidrule(lr){7-9}
%  \midrule
  $n_{\text{pess}}$ & $\sigma^{2}_{\beta}$ & $p$ & $\beta_{1}$ & $\beta_{2}$ & $\beta_{3}$ & $\beta_{1}$ & $\beta_{2}$ & $\beta_{3}$ \\ 
  \midrule
30 & 5 & 0.10 & 230.77 & 0.41 & 0.15 & 152.72 & 0.31 & 0.16 \\ 
   &  & 0.25 & 3.50 & 0.41 & 0.11 & 5.44 & 0.16 & 0.10 \\ 
   &  & 0.50 & 0.05 & 0.41 & 0.11 & 0.02 & 0.14 & 0.05 \\ 
   &  & 0.75 & 1.10 & 0.38 & 0.11 & 4.96 & 0.20 & 0.09 \\ 
   &  & 0.90 & 93.94 & 0.31 & 0.13 & 141.53 & 0.32 & 0.15 \\ 
    \midrule
%30 & 10 & 0.10 & 243.68 & 0.45 & 0.16 & 206.88 & 0.42 & 0.23 \\ 
%   &  & 0.25 & 3.74 & 0.43 & 0.11 & 6.03 & 0.25 & 0.12 \\ 
%   &  & 0.50 & 0.06 & 0.41 & 0.11 & 0.04 & 0.13 & 0.05 \\ 
%   &  & 0.75 & 1.22 & 0.35 & 0.09 & 6.59 & 0.38 & 0.16 \\ 
%   &  & 0.90 & 99.69 & 0.30 & 0.16 & 202.90 & 0.20 & 0.26 \\
%    \midrule
50 & 5 & 0.10 & 225.28 & 0.43 & 0.15 & 146.22 & 0.29 & 0.19 \\ 
   &  & 0.25 & 4.76 & 0.37 & 0.11 & 7.20 & 0.17 & 0.11 \\ 
   &  & 0.50 & 0.05 & 0.39 & 0.10 & 0.02 & 0.12 & 0.04 \\ 
   &  & 0.75 & 1.30 & 0.35 & 0.10 & 6.11 & 0.23 & 0.11 \\ 
   &  & 0.90 & 97.72 & 0.34 & 0.12 & 147.15 & 0.26 & 0.14 \\
   \midrule
50 & 10 & 0.10 & 217.57 & 0.47 & 0.16 & 193.53 & 0.25 & 0.22 \\ 
   &  & 0.25 & 3.73 & 0.40 & 0.12 & 6.73 & 0.27 & 0.18 \\ 
   &  & 0.50 & 0.06 & 0.39 & 0.10 & 0.04 & 0.13 & 0.04 \\ 
   &  & 0.75 & 1.22 & 0.35 & 0.09 & 6.06 & 0.32 & 0.16 \\ 
   &  & 0.90 & 106.17 & 0.32 & 0.16 & 206.09 & 0.32 & 0.25 \\ 
    \midrule
100 & 5 & 0.10 & 261.97 & 0.35 & 0.18 & 181.99 & 0.20 & 0.16 \\ 
   &  & 0.25 & 4.76 & 0.36 & 0.11 & 6.92 & 0.17 & 0.09 \\ 
   &  & 0.50 & 0.05 & 0.41 & 0.11 & 0.02 & 0.13 & 0.05 \\ 
   &  & 0.75 & 1.25 & 0.36 & 0.10 & 5.51 & 0.21 & 0.11 \\ 
   &  & 0.90 & 92.45 & 0.29 & 0.14 & 153.69 & 0.18 & 0.16 \\
    \midrule
100 & 10 & 0.10 & 275.68 & 0.40 & 0.14 & 258.67 & 0.39 & 0.26 \\ 
   &  & 0.25 & 4.12 & 0.38 & 0.11 & 6.89 & 0.18 & 0.10 \\ 
   &  & 0.50 & 0.05 & 0.38 & 0.10 & 0.03 & 0.11 & 0.04 \\ 
   &  & 0.75 & 1.39 & 0.37 & 0.11 & 5.80 & 0.24 & 0.11 \\ 
   &  & 0.90 & 95.36 & 0.34 & 0.15 & 209.48 & 0.38 & 0.29 \\
%    \midrule
%100 & 10 & 0.10 & 133.62 & 0.28 & 0.15 & 264.36 & 0.32 & 0.26 \\ 
%   &  & 0.25 & 3.13 & 0.20 & 0.07 & 9.83 & 0.63 & 0.25 \\ 
%   &  & 0.50 & 0.07 & 0.20 & 0.06 & 0.03 & 0.09 & 0.03 \\ 
%%   &  & 0.75 & 0.58 & 0.16 & 0.05 & 6.51 & 0.64 & 0.25 \\ 
%   &  & 0.90 & 40.92 & 0.22 & 0.09 & 241.34 & 0.71 & 0.31 \\
   \bottomrule
\end{tabular}
\end{table}
%-------------------------------------------------------------------------------

%-------------------------------------------------------------------------------
\begin{table}[ht]
\centering
\caption{Coverage rates of posterior estimates of regression coefficients from the two models when $\delta_{01}=0.3$ and $\delta_{10}=0.2$ under different settings of prior variance and prior effective sample size ($n_{\text{pess}}$).} 
\label{Table:Case2_Coverage}
\begin{tabular}{ccc ccc ccc}
\toprule
 &  &  & \multicolumn{3}{c}{BQR} & \multicolumn{3}{c}{BQRM} \\
\cmidrule(lr){4-6} \cmidrule(lr){7-9}
%  \midrule
  $n_{\text{pess}}$ & $\sigma^{2}_{\beta}$ & $p$ & $\beta_{1}$ & $\beta_{2}$ & $\beta_{3}$ & $\beta_{1}$ & $\beta_{2}$ & $\beta_{3}$ \\ 
  \midrule
30 & 5 & 0.10 & 0.25 & 0.61 & 0.60 & 1.00 & 1.00 & 1.00 \\ 
   &  & 0.25 & 0.06 & 0.44 & 0.43 & 1.00 & 1.00 & 1.00 \\ 
   &  & 0.50 & 0.22 & 0.03 & 0.31 & 1.00 & 0.99 & 1.00 \\ 
   &  & 0.75 & 0.91 & 0.48 & 0.51 & 1.00 & 1.00 & 1.00 \\ 
   &  & 0.90 & 0.23 & 0.76 & 0.69 & 1.00 & 1.00 & 1.00 \\ 
    \midrule
%30 & 10 & 0.10 & 0.25 & 0.56 & 0.57 & 1.00 & 1.00 & 1.00 \\ 
 %  &  & 0.25 & 0.02 & 0.37 & 0.40 & 1.00 & 1.00 & 1.00 \\ 
 %  &  & 0.50 & 0.20 & 0.06 & 0.37 & 1.00 & 0.99 & 0.97 \\ 
 %  &  & 0.75 & 0.91 & 0.56 & 0.62 & 1.00 & 1.00 & 1.00 \\ 
 %  &  & 0.90 & 0.24 & 0.73 & 0.71 & 1.00 & 1.00 & 1.00 \\ 
  %  \midrule
50 & 5 & 0.10 & 0.29 & 0.59 & 0.57 & 1.00 & 1.00 & 1.00 \\ 
   &  & 0.25 & 0.05 & 0.48 & 0.47 & 1.00 & 1.00 & 1.00 \\ 
   &  & 0.50 & 0.13 & 0.07 & 0.40 & 1.00 & 1.00 & 0.99 \\ 
   &  & 0.75 & 0.90 & 0.55 & 0.54 & 1.00 & 1.00 & 0.98 \\ 
   &  & 0.90 & 0.26 & 0.70 & 0.70 & 1.00 & 1.00 & 1.00 \\ 
   \midrule
50 & 10 & 0.10 & 0.30 & 0.54 & 0.56 & 1.00 & 1.00 & 1.00 \\ 
   &  & 0.25 & 0.02 & 0.41 & 0.44 & 1.00 & 1.00 & 1.00 \\ 
   &  & 0.50 & 0.21 & 0.06 & 0.34 & 1.00 & 0.99 & 0.99 \\ 
   &  & 0.75 & 0.86 & 0.54 & 0.60 & 1.00 & 1.00 & 1.00 \\ 
   &  & 0.90 & 0.27 & 0.70 & 0.65 & 1.00 & 1.00 & 1.00 \\ 
   \midrule
100 & 5 & 0.10 & 0.17 & 0.66 & 0.67 & 1.00 & 1.00 & 1.00 \\ 
   &  & 0.25 & 0.07 & 0.48 & 0.50 & 1.00 & 1.00 & 1.00 \\ 
   &  & 0.50 & 0.21 & 0.04 & 0.30 & 1.00 & 1.00 & 0.96 \\ 
   &  & 0.75 & 0.88 & 0.52 & 0.56 & 1.00 & 1.00 & 1.00 \\ 
   &  & 0.90 & 0.27 & 0.73 & 0.66 & 1.00 & 1.00 & 1.00 \\ 
   \midrule
100 & 10 & 0.10 & 0.14 & 0.63 & 0.67 & 1.00 & 1.00 & 1.00 \\ 
   &  & 0.25 & 0.06 & 0.45 & 0.50 & 1.00 & 1.00 & 1.00 \\ 
   &  & 0.50 & 0.24 & 0.09 & 0.36 & 1.00 & 1.00 & 0.97 \\ 
   &  & 0.75 & 0.80 & 0.50 & 0.52 & 1.00 & 1.00 & 1.00 \\ 
   &  & 0.90 & 0.30 & 0.67 & 0.65 & 1.00 & 1.00 & 1.00 \\ 
   \bottomrule
\end{tabular}
\end{table}
%-------------------------------------------------------------------------------

%-------------------------------------------------------------------------------
%                           Third
%-------------------------------------------------------------------------------

%-------------------------------------------------------------------------------
\begin{table}[ht]
\centering
\caption{Average biases of posterior estimates of regression coefficients from the two models when $\delta_{01}=0.2$ and $\delta_{10}=0.1$ under different settings of prior variance and prior effective sample size ($n_{\text{pess}}$).} 
\label{Table:Case3_AB}
\begin{tabular}{ccc ccc ccc}
\toprule
 &  &  & \multicolumn{3}{c}{BQR} & \multicolumn{3}{c}{BQRM} \\
\cmidrule(lr){4-6} \cmidrule(lr){7-9}
%  \midrule
  $n_{\text{pess}}$ & $\sigma^{2}_{\beta}$ & $p$ & $\beta_{1}$ & $\beta_{2}$ & $\beta_{3}$ & $\beta_{1}$ & $\beta_{2}$ & $\beta_{3}$ \\ 
  \midrule
30 & 5 & 0.10 & -9.11 & -0.20 & 0.12 & -11.55 & 0.04 & 0.05 \\ 
   &  & 0.25 & -1.31 & -0.36 & 0.20 & -2.06 & 0.57 & -0.23 \\ 
   &  & 0.50 & -0.24 & -0.45 & 0.20 & -0.01 & -0.07 & 0.00 \\ 
   &  & 0.75 & 0.52 & -0.35 & 0.17 & 1.84 & 0.61 & -0.30 \\ 
   &  & 0.90 & 5.25 & -0.14 & 0.06 & 11.12 & 0.24 & -0.15 \\ 
    \midrule
%30 & 10 & 0.10 & -9.46 & -0.21 & 0.10 & -13.37 & 0.14 & -0.09 \\ 
%   &  & 0.25 & -1.45 & -0.36 & 0.18 & -2.36 & 0.64 & -0.33 \\ 
%   &  & 0.50 & -0.24 & -0.44 & 0.21 & 0.00 & -0.10 & 0.04 \\ 
%   &  & 0.75 & 0.48 & -0.37 & 0.18 & 1.80 & 0.67 & -0.34 \\ 
%   &  & 0.90 & 6.19 & 0.01 & -0.01 & 15.07 & 0.43 & -0.29 \\
%    \midrule
50 & 5 & 0.10 & -9.87 & -0.16 & 0.07 & -12.15 & 0.04 & -0.07 \\ 
   &  & 0.25 & -1.31 & -0.38 & 0.17 & -1.97 & 0.47 & -0.28 \\ 
   &  & 0.50 & -0.24 & -0.44 & 0.21 & -0.01 & -0.11 & 0.03 \\ 
   &  & 0.75 & 0.52 & -0.36 & 0.18 & 1.83 & 0.58 & -0.27 \\ 
   &  & 0.90 & 5.84 & -0.06 & 0.04 & 12.10 & 0.21 & -0.12 \\
    \midrule
50 & 10 & 0.10 & -8.41 & -0.33 & 0.11 & -12.57 & 0.19 & -0.12 \\ 
   &  & 0.25 & -1.35 & -0.37 & 0.19 & -2.21 & 0.57 & -0.27 \\ 
   &  & 0.50 & -0.24 & -0.43 & 0.22 & -0.03 & -0.09 & 0.05 \\ 
   &  & 0.75 & 0.51 & -0.35 & 0.19 & 1.86 & 0.71 & -0.29 \\ 
   &  & 0.90 & 5.82 & -0.08 & -0.01 & 13.37 & 0.38 & -0.33 \\ 
   \midrule
100 & 5 & 0.10 & -8.53 & -0.30 & 0.06 & -11.05 & 0.01 & -0.13 \\ 
   &  & 0.25 & -1.38 & -0.38 & 0.19 & -2.02 & 0.44 & -0.21 \\ 
   &  & 0.50 & -0.24 & -0.43 & 0.23 & -0.01 & -0.10 & 0.07 \\ 
   &  & 0.75 & 0.57 & -0.34 & 0.18 & 2.06 & 0.58 & -0.26 \\ 
   &  & 0.90 & 5.64 & -0.09 & 0.04 & 11.76 & 0.30 & -0.14 \\
   \midrule
   100 & 10 & 0.10 & -9.53 & -0.24 & 0.08 & -13.91 & 0.19 & -0.16 \\ 
   &  & 0.25 & -1.46 & -0.36 & 0.17 & -2.32 & 0.60 & -0.31 \\ 
   &  & 0.50 & -0.24 & -0.43 & 0.23 & -0.03 & -0.09 & 0.07 \\ 
   &  & 0.75 & 0.57 & -0.35 & 0.17 & 1.95 & 0.60 & -0.33 \\ 
   &  & 0.90 & 5.49 & -0.07 & 0.05 & 13.66 & 0.56 & -0.25 \\
   \bottomrule
\end{tabular}
\end{table}
%-------------------------------------------------------------------------------

%-------------------------------------------------------------------------------
\begin{table}[ht]
\centering
\caption{Mean squared errors (MSE) of posterior estimates of regression coefficients from the two models when $\delta_{01}=0.2$ and $\delta_{10}=0.1$ under different settings of prior variance and prior effective sample size ($n_{\text{pess}}$).} 
\label{Table:Case3_MSE}
\begin{tabular}{ccc ccc ccc}
\toprule
 &  &  & \multicolumn{3}{c}{BQR} & \multicolumn{3}{c}{BQRM} \\
\cmidrule(lr){4-6} \cmidrule(lr){7-9}
%  \midrule
  $n_{\text{pess}}$ & $\sigma^{2}_{\beta}$ & $p$ & $\beta_{1}$ & $\beta_{2}$ & $\beta_{3}$ & $\beta_{1}$ & $\beta_{2}$ & $\beta_{3}$ \\ 
  \midrule
30 & 5 & 0.10 & 115.73 & 0.27 & 0.12 & 178.40 & 0.18 & 0.12 \\ 
   &  & 0.25 & 2.50 & 0.21 & 0.07 & 8.47 & 0.55 & 0.16 \\ 
   &  & 0.50 & 0.07 & 0.22 & 0.05 & 0.04 & 0.16 & 0.05 \\ 
   &  & 0.75 & 0.55 & 0.17 & 0.06 & 6.58 & 0.59 & 0.23 \\ 
   &  & 0.90 & 35.43 & 0.17 & 0.09 & 150.43 & 0.22 & 0.15 \\ 
    \midrule
%30 & 10 & 0.10 & 129.14 & 0.29 & 0.13 & 241.20 & 0.28 & 0.21 \\ 
%   &  & 0.25 & 3.06 & 0.20 & 0.06 & 9.88 & 0.68 & 0.25 \\ 
%   &  & 0.50 & 0.06 & 0.21 & 0.05 & 0.07 & 0.10 & 0.03 \\ 
%   &  & 0.75 & 0.42 & 0.18 & 0.05 & 6.22 & 0.77 & 0.25 \\ 
%   &  & 0.90 & 47.16 & 0.16 & 0.11 & 256.05 & 0.43 & 0.35 \\
%    \midrule
50 & 5 & 0.10 & 135.37 & 0.29 & 0.12 & 195.33 & 0.20 & 0.14 \\ 
   &  & 0.25 & 2.64 & 0.23 & 0.06 & 8.05 & 0.48 & 0.19 \\ 
   &  & 0.50 & 0.06 & 0.21 & 0.05 & 0.03 & 0.11 & 0.03 \\ 
   &  & 0.75 & 0.49 & 0.16 & 0.06 & 5.85 & 0.54 & 0.21 \\ 
   &  & 0.90 & 42.36 & 0.17 & 0.13 & 169.76 & 0.16 & 0.20 \\
   \midrule
50 & 10 & 0.10 & 113.37 & 0.31 & 0.12 & 235.78 & 0.31 & 0.23 \\ 
   &  & 0.25 & 2.60 & 0.21 & 0.06 & 9.14 & 0.62 & 0.22 \\ 
   &  & 0.50 & 0.06 & 0.20 & 0.05 & 0.05 & 0.11 & 0.02 \\ 
   &  & 0.75 & 0.51 & 0.18 & 0.06 & 6.67 & 0.87 & 0.23 \\ 
   &  & 0.90 & 45.80 & 0.22 & 0.10 & 230.12 & 0.42 & 0.35 \\
    \midrule
100 & 5 & 0.10 & 112.56 & 0.32 & 0.13 & 180.96 & 0.22 & 0.19 \\ 
   &  & 0.25 & 2.89 & 0.22 & 0.06 & 8.07 & 0.44 & 0.15 \\ 
   &  & 0.50 & 0.06 & 0.21 & 0.06 & 0.03 & 0.13 & 0.03 \\ 
   &  & 0.75 & 0.56 & 0.15 & 0.05 & 6.93 & 0.56 & 0.19 \\ 
   &  & 0.90 & 41.72 & 0.19 & 0.11 & 170.27 & 0.25 & 0.15 \\
   \midrule
100 & 10 & 0.10 & 133.62 & 0.28 & 0.15 & 264.36 & 0.32 & 0.26 \\ 
   &  & 0.25 & 3.13 & 0.20 & 0.07 & 9.83 & 0.63 & 0.25 \\ 
   &  & 0.50 & 0.07 & 0.20 & 0.06 & 0.03 & 0.09 & 0.03 \\ 
   &  & 0.75 & 0.58 & 0.16 & 0.05 & 6.51 & 0.64 & 0.25 \\ 
   &  & 0.90 & 40.92 & 0.22 & 0.09 & 241.34 & 0.71 & 0.31 \\ 
   \bottomrule
\end{tabular}
\end{table}
%-------------------------------------------------------------------------------

%-------------------------------------------------------------------------------
\begin{table}[ht]
\centering
\caption{Coverage rates of posterior estimates of regression coefficients from the two models when $\delta_{01}=0.2$ and $\delta_{10}=0.1$ under different settings of prior variance and prior effective sample size ($n_{\text{pess}}$).}
\label{Table:Case3_Coverage}
\begin{tabular}{ccc ccc ccc}
\toprule
 &  &  & \multicolumn{3}{c}{BQR} & \multicolumn{3}{c}{BQRM} \\
\cmidrule(lr){4-6} \cmidrule(lr){7-9}
%  \midrule
  $n_{\text{pess}}$ & $\sigma^{2}_{\beta}$ & $p$ & $\beta_{1}$ & $\beta_{2}$ & $\beta_{3}$ & $\beta_{1}$ & $\beta_{2}$ & $\beta_{3}$ \\ 
  \midrule
30 & 5 & 0.10 & 0.26 & 0.73 & 0.72 & 1.00 & 1.00 & 1.00 \\ 
   &  & 0.25 & 0.37 & 0.64 & 0.64 & 1.00 & 1.00 & 1.00 \\ 
   &  & 0.50 & 0.23 & 0.59 & 0.83 & 1.00 & 0.99 & 1.00 \\ 
   &  & 0.75 & 0.98 & 0.71 & 0.75 & 0.99 & 1.00 & 1.00 \\ 
   &  & 0.90 & 0.90 & 0.83 & 0.83 & 1.00 & 1.00 & 1.00 \\ 
    \midrule
%30 & 10 & 0.10 & 0.29 & 0.72 & 0.68 & 0.99 & 1.00 & 1.00 \\ 
%   &  & 0.25 & 0.34 & 0.67 & 0.65 & 1.00 & 1.00 & 1.00 \\ 
%   &  & 0.50 & 0.17 & 0.60 & 0.82 & 1.00 & 1.00 & 1.00 \\ 
%   &  & 0.75 & 1.00 & 0.68 & 0.76 & 1.00 & 1.00 & 0.99 \\ 
%   &  & 0.90 & 0.77 & 0.89 & 0.85 & 1.00 & 1.00 & 1.00 \\ 
%   \midrule
50 & 5 & 0.10 & 0.25 & 0.75 & 0.74 & 1.00 & 1.00 & 1.00 \\ 
   &  & 0.25 & 0.43 & 0.61 & 0.69 & 1.00 & 0.99 & 1.00 \\ 
   &  & 0.50 & 0.19 & 0.56 & 0.81 & 0.98 & 1.00 & 1.00 \\ 
   &  & 0.75 & 0.99 & 0.73 & 0.74 & 1.00 & 1.00 & 1.00 \\ 
   &  & 0.90 & 0.81 & 0.88 & 0.84 & 1.00 & 1.00 & 1.00 \\
    \midrule
50 & 10 & 0.10 & 0.35 & 0.65 & 0.65 & 1.00 & 1.00 & 1.00 \\ 
   &  & 0.25 & 0.39 & 0.66 & 0.68 & 1.00 & 1.00 & 0.98 \\ 
   &  & 0.50 & 0.20 & 0.63 & 0.83 & 1.00 & 1.00 & 1.00 \\ 
   &  & 0.75 & 0.98 & 0.70 & 0.70 & 0.98 & 1.00 & 1.00 \\ 
   &  & 0.90 & 0.80 & 0.81 & 0.81 & 1.00 & 1.00 & 1.00 \\ 
   \midrule
100 & 5 & 0.10 & 0.33 & 0.66 & 0.67 & 1.00 & 1.00 & 1.00 \\ 
   &  & 0.25 & 0.37 & 0.62 & 0.64 & 1.00 & 1.00 & 0.99 \\ 
   &  & 0.50 & 0.18 & 0.60 & 0.74 & 1.00 & 1.00 & 0.97 \\ 
   &  & 0.75 & 0.98 & 0.76 & 0.78 & 1.00 & 1.00 & 1.00 \\ 
   &  & 0.90 & 0.84 & 0.84 & 0.81 & 1.00 & 1.00 & 1.00 \\
         \midrule
  100 & 10 & 0.10 & 0.29 & 0.72 & 0.72 & 0.99 & 1.00 & 1.00 \\ 
   &  & 0.25 & 0.35 & 0.68 & 0.67 & 0.99 & 1.00 & 1.00 \\ 
   &  & 0.50 & 0.13 & 0.69 & 0.71 & 0.98 & 1.00 & 0.99 \\ 
   &  & 0.75 & 0.97 & 0.72 & 0.79 & 0.99 & 1.00 & 1.00 \\ 
   &  & 0.90 & 0.82 & 0.80 & 0.79 & 1.00 & 1.00 & 1.00 \\
   \bottomrule
\end{tabular}
\end{table}
%-------------------------------------------------------------------------------

\clearpage
\newpage
\section{Additional Analysis of India DHS Data}
\label{app:india}

\begin{table}[ht]
\centering
\begin{tabular}{l l S S S S S S}
\toprule
& & \multicolumn{3}{c}{BQR}
    & \multicolumn{3}{c}{BQRM} \\
\cmidrule(lr){3-5}\cmidrule(lr){6-8}

Quantile & Variable
& \multicolumn{1}{c}{Mean}
& \multicolumn{2}{c}{95\% CrI}
& \multicolumn{1}{c}{Mean}
& \multicolumn{2}{c}{95\% CrI} \\
\cmidrule(lr){4-5}\cmidrule(lr){7-8}

& & & \multicolumn{1}{c}{Lower} & \multicolumn{1}{c}{Upper}
& & \multicolumn{1}{c}{Lower} & \multicolumn{1}{c}{Upper} \\
\midrule

0.1 & Intercept  & -5.21 & -9.48 & -2.71 & -6.78  & -51.79 & 5.25 \\
    & fage       & -0.37 & -0.63 & -0.20 & -0.95  & -5.36  & 0.24 \\
    & fwork      &  0.29 &  0.09 &  0.56 &  0.83  & -0.94  & 5.22 \\
    & meduc      & -0.47 & -0.82 & -0.26 & -1.33  & -6.59  & 0.14 \\
    & wealth     & -1.02 & -1.68 & -0.62 & -3.38  & -15.65 & -0.16 \\
    & nchildren  &  0.34 &  0.20 &  0.58 &  1.29  & -0.02  & 5.66 \\
    & remarriage &  0.29 & -0.24 &  0.91 &  1.55  & -2.53  & 7.89 \\
    & polyg      &  1.32 &  0.59 &  2.34 &  4.54  & -0.61  & 16.31 \\
    & nwomen     & -0.07  & -0.17 &  0.02 & -0.17  & -1.70  & 0.65 \\
    & $\delta_{01}$ & {\text{--}} & {\text{--}} & {\text{--}}
                      & 0.70 & 0.07 & 0.83 \\
    & $\delta_{10}$ & {\text{--}} & {\text{--}} & {\text{--}}
                      & 0.03 & 0.00 & 0.16 \\

\midrule

0.9 & Intercept  &  4.97  & -0.41 & 22.02 & 21.60  & -1.61  & 67.28 \\
    & fage       & -0.56  & -1.75 & -0.11 & -0.93  & -4.86  & 0.32 \\
    & fwork      &  0.55  &  0.09 &  1.63 &  1.45  & -0.42  & 8.32 \\
    & meduc      & -0.67  & -2.19 & -0.12 & -0.95  & -5.13  & 0.58 \\
    & wealth     & -1.61  & -4.93 & -0.30 & -2.39  & -10.57 & -0.09 \\
    & nchildren  &  0.56  &  0.10 &  1.69 &  0.88  & -0.25  & 4.17 \\
    & remarriage &  0.58  & -0.22 &  2.08 &  0.95  & -4.62  & 7.97 \\
    & polyg      &  2.50  &  0.45 &  6.71 &  3.23  & -2.58  & 12.99 \\
    & nwomen     & -0.11  & -0.39 &  0.01 & -0.23  & -2.06  & 0.99 \\
    & $\delta_{01}$ & {\text{--}} & {\text{--}} & {\text{--}}
                      & 0.62 & 0.02 & 0.83 \\
    & $\delta_{10}$ & {\text{--}} & {\text{--}} & {\text{--}}
                      & 0.03 & 0.00 & 0.12 \\

\bottomrule
\end{tabular}

\caption{Posterior means and 95\% credible intervals (CrI).}
\label{Table:AppPostResSupp}
\end{table}

\end{document}